\documentclass[preprint,aps,nofootinbib]{revtex4-1}

\usepackage{dcolumn}
\usepackage{bm}
\usepackage{epsfig}
\usepackage{graphicx,epstopdf}
\usepackage{float}
\usepackage{paralist}
\usepackage{comment}

\usepackage{amsmath}
\usepackage{amsfonts}
\usepackage{amssymb}
\usepackage{adjustbox}
\usepackage{color}
\usepackage{xcolor}
\usepackage{adjustbox}
\usepackage{subcaption}
\usepackage[toc,page]{appendix}
\usepackage[utf8]{inputenc}
\usepackage{slashed}
\usepackage{indentfirst}
\usepackage[justification=justified,singlelinecheck=false]{caption}
\usepackage{subcaption}
\usepackage{hyperref} 
\usepackage{natbib} 
\usepackage{ulem}
\def\tt{\tilde{t}}
\def\lsp{\tilde{\chi}^0_1}
\def\mtt{m_{\tilde{t}}}
\def\mchar{m_{\tilde{\chi}_1^{\pm}}}
\def\mneutwo{m_{\tilde{\chi}_2^{0}}}

\def\mlsp{m_{\tilde{\chi}_1^{0}}}
\def\Rmb{\bar{R}_M}
\def\Rmbmax{\bar{R}_{\text{max}}}
\def\Rmbmin{\bar{R}_{\text{min}}}
\def\ifb{\text{fb}^{-1}}
\def\nlsp{\tilde{\chi}^0_2}

\begin{document}

\title{
Constraining the Compressed Top Squark and Chargino along the W Corridor}
\author{Hsin-Chia Cheng,$^{a,b}$ Lingfeng Li$^a$, Qin Qin$^a$}
\email[Email: ]{cheng@physics.ucdavis.edu}\email{llfli@ucdavis.edu}\email{qqin@ucdavis.edu}
\affiliation{$^a$Department of Physics, University of California Davis, Davis, California 95616, USA\\
$^b$School of Natural Sciences, Institute for Advanced Study, Princeton, New Jersey 08540, USA}

\begin{abstract}
Studying superpartner production together with a hard initial state radiation (ISR) jet has been a useful strategy for searches of supersymmetry with a compressed spectrum at the Large Hadron Collider (LHC). In the case of the top squark (stop), the ratio of the missing transverse momentum from the lightest neutralinos and the ISR momentum, defined as $\bar{R}_M$, turns out to be an effective variable to distinguish the signal from the backgrounds. It has helped to exclude the stop mass below 590 GeV along the top corridor where $m_{\tilde{t}} - m_{\tilde{\chi}_1^0} \approx m_t$. On the other hand, the current experimental limit is still rather weak in the $W$ corridor where $m_{\tilde{t}} - m_{\tilde{\chi}_1^0} \approx m_W +m_b$. In this work we extend this strategy to the parameter region around the $W$ corridor by considering the one lepton final state. In this case the kinematic constraints are insufficient to completely determine the neutrino momentum which is required to calculate $\bar{R}_M$. However, the minimum value of $\bar{R}_M$ consistent with the kinematic constraints still provides a useful discriminating variable, allowing the exclusion reach of the stop mass to be extended to $\sim 550$ GeV based on the current 36 fb$^{-1}$ LHC data. The same method can also be applied to the chargino search with $m_{\tilde{\chi}_1^\pm} -m_{\tilde{\chi}_1^0} \approx m_W$ because the analysis does not rely on $b$ jets. If no excess is present in the current data, a chargino mass of 300 GeV along the $W$ corridor can be excluded, beyond the limit obtained from the multilepton search.

\end{abstract}
\maketitle

\section{Introduction}

Weak-scale supersymmetry (SUSY) has long been considered as the leading candidate for the new physics beyond the Standard Model (SM). Supersymmetric particles have been searched for at colliders for decades but unfortunately none of them has been found yet. The strong limits from the searches at the Large Hadron Collider (LHC) have raised concerns if SUSY can provide the solution to the hierarchy problem of the Higgs mass scale in SM, as the LHC probes the energy scale into the TeV range. For the hierarchy problem, the most relevant particles are the superpartners of the top quark, top squarks (or simply stops), because the top quark has the largest coupling to the Higgs field and hence gives the largest quadratic correction to the Higgs mass-squared parameter. The stops are needed to be near the weak scale to cut off this contribution in order for the theory to be natural. The current lower bound on the stop has reached beyond 1 TeV in typical search channels at the LHC~\cite{Aaboud:2017ayj,Aaboud:2017bac,Aaboud:2017nfd,Sirunyan:2017cwe,Sirunyan:2017wif,Sirunyan:2017kqq,CMS:2017zki,Sirunyan:2017xse,Sirunyan:2017leh}, which would imply quite severe fine-tuning already.

Of course, there are cases where the stop mass limit is not as strong yet. In particular, the limit degrades if the masses difference between the stop and the lightest supersymmetric particle (LSP) that it decays to become smaller, i.e., they have a compressed spectrum. In this case, the visible SM particles from the stop decay will not carry a large amount of energy. The missing transverse momentum will also be suppressed because it is just the opposite of the sum of the visible transverse momentum. These signal events are more difficult to be distinguished from the SM backgrounds. The search strategies and search limits depend on how compressed the spectrum is and stop decay chains. For example, for highly compressed spectra ($\mtt-\mlsp < m_W$), the searches rely on 4-body decays  {($\tt \to b\lsp ff'$)}~\cite{Boehm:1999tr,Das:2001kd,Konar:2016ata,Aaboud:2017bac,Aaboud:2017nfd,CMS:2017odo,CMS:2016zvj,Khachatryan:2016pxa} or flavor-changing decays {($\tt \to c\lsp W$)}~\cite{Hikasa:1987db,Muhlleitner:2011ww,Khachatryan:2016pxa,Sirunyan:2017kiw,Aad:2014nra,Aaboud:2016tnv}, or even monojet~\cite{Drees:2012dd}.  Before the LHC Run 2, the most difficult case used to be the top corridor, where $\mtt \approx m_t + \mlsp$. This is because the top quark and the neutralino from the stop decay carry little momenta in the stop rest frame and are boosted with the same velocity as the original stop particle. The stop pair is produced back-to-back in the transverse plane, resulting in the cancellation of the two neutralinos' transverse momenta, leaving no trace of their existence. Then the events look exactly like the large SM $t\bar{t}$ background. The analyses of the Run 1 data provided essentially no constraint along the top corridor. 

The situation has completely changed in Run 2 with new techniques being employed to attack this parameter space region.  An important observation is that if the stop pair is produced with a hard initial state radiation (ISR) jet, the two neutralinos will be boosted in the opposite direction to the ISR jet, giving missing transverse energy (MET) anti-parallel to the ISR jet~\cite{Carena:2008mj,Hagiwara:2013tva,An:2015uwa,Macaluso:2015wja}. A variable 
$R_M$ which measures the ratio the two-neutralino transverse momentum and the ISR transverse momentum provides a powerful discriminator between the signal and the backgrounds, as it  should be equal to $\mlsp/\mtt$ for the stop events while close to zero for the SM $t\bar{t}$ backgrounds. With a sophisticated method to determine the ISR system~\cite{Jackson:2016mfb}, the Run 2 analysis has been able to exclude the stop mass below 590 GeV along the top corridor with 36 fb$^{-1}$ of data~\cite{Aaboud:2017ayj}, assuming 100\% branching fraction to $t\lsp$. This is quite impressive, and the limit is even stronger than the nearby parameter region where the mass difference between $\tilde{t}$ and $\lsp$ is somewhat off the top quark mass.

The analysis with the $R_M$ variable is based on the fully hadronic channel where there is no additional MET other than that carried by the two neutralinos. In this case $R_M$ is simply given by
\begin{equation}\label{RM}
R_M\equiv\frac{\slashed{p}_T}{p_{T(J_{\text{ISR}})}}\approx \frac{m_{\tilde{\chi}_1^0}}{\mtt}\ .
\end{equation} 
In semileptonic or dileptonic decays, however, the neutrino(s) coming from the $W$ decays give an additional contribution to MET, which ruin the relation in Eq.~(\ref{RM}). One way to deal with this is to try to reconstruct the neutrino momentum so that it can be subtracted from the total MET to obtain the MET due to the neutralinos only. Then a modified $\bar{R}_M$ variable related to $\mlsp/\mtt$ can be defined analogously as a discriminating variable. For the semileptonic decay, it was shown~\cite{Cheng:2016mcw} that from the 3 mass shell conditions ($m_t,\, m_W, \, m_\nu $) together with the assumption that the perpendicular component of the $\slashed{p}_T$ relative to $p_{T(J_{\text{ISR}})}$ is entirely due to the neutrino, the neutrino momentum can be solved with a two-fold ambiguity. After subtracting the solved neutrino momentum, the $\bar{R}_M$ variable also provides a strong discriminator for the stop events in the semileptonic decay channel and makes it competitive with the fully hadronic result.

For the dileptonic channel with two final state neutrinos, there is one more unknown than the number of kinematic constraint equations, so we can not completely reconstruct the neutrino momenta. Instead, for each event we can only obtain a finite range of $\bar{R}_M$ which can be consistent with that event. Nevertheless, the upper and lower limits of the allowed $\bar{R}_M$, denoted by $\Rmbmax$ and $\Rmbmin$ could provide potential variables for discriminating signals from backgrounds. In Ref.~\cite{Cheng:2017dxe}, it was found indeed that 
$\Rmbmax$ and $\Rmbmin$ provide more discriminating power than just using $p_{T(J_{\text{ISR}})}$ and MET. Although, for the case of stop decaying to top plus LSP, the dileptonic channel is not expected to compete with the all-hadronic or semileptonic channels due to the small branching ratios, the dileptonic search can be useful if the SUSY spectrum is such that the stop decays mainly through the chargino and the slepton decays to the LSP, in which case the dileptonic final states can be dominant~\cite{Cheng:2017dxe,Konar:2017oah}. 

After the progress in stop search coverage along the top corridor, the $W$ corridor where $\mtt \approx m_W+ m_b+ \mlsp$ remains relatively weakly constrained. In this case, the bottom quark, $W$ and $\lsp$ from the stop decay are also static in the stop rest frame. The missing $p_T$ from the two $\lsp$'s again cancels from the back-to-back boost of the stop-pair in the transverse plane. Such events are difficult to be distinguished from the SM backgrounds, resulting a poor reach in current LHC searches and the stop could still be as light as $\sim 360$ GeV around that region~\cite{Aaboud:2017bac,Aaboud:2017nfd}. A natural thought is again to consider events with an ISR jet to boost the stop-pair in the opposite direction so that the $\lsp$'s will produce some MET. Then one can use the similar $R_M$ variables to distinguish signals from backgrounds. A main goal of this study is to explore whether this technique can help to improve the stop mass bound around the $W$ corridor.

We will focus on the semileptonic events where one $W$ decays leptonically and the other $W$ decays hadronically. The $b$-jets from the stop decays will be soft so they will not be useful due to low tagging efficiencies and large hadronic backgrounds. Compared with the semileptonic stop events along the top corridor, we lose a top quark mass shell constraint because the decay does not go through an on-shell top quark. Therefore the neutrino momentum can not be completely reconstructed and a unique $\Rmb$ value can not be obtained. Nevertheless, the kinematic constraints still limit $\Rmb$ into a finite range. We can define the $\Rmbmax$ and $\Rmbmin$ variables just as for the case of the dileptonic stop events in the top corridor to examine whether they are useful in suppressing backgrounds. We will find out that $\Rmbmin$ does provide a useful discriminating variable in this case.

Since the $b$-jets are too soft to be useful in the $W$ corridor of the stop, the signal events look the same as the chargino pair production in the $W$ corridor ($\mchar \approx m_W + \mlsp$) if one ignores the $b$-jets. The same analysis can be applied to the chargino search around the $W$ corridor. In SUSY, the chargino is usually accompanied by one or two neutralinos with similar masses, depending on whether it is wino-like or higgsino-like. Hence, one should consider chargino and neutralino pair productions altogether.  
Under the assumption that the LSP is bino-like and the chargino (neutralino) decays to the LSP plus an (one-shell or off-shell) $W$ ($Z$), 
the current strongest constraints come from tri-lepton searches~\cite{Sirunyan:2017lae,CMS:2017sqn,Aad:2014nua,Aad:2014vma,ATLAS:2017uun}, where the production cross section is taken to be coming from the winos. However, for the same reason due to the large SM $WZ$ background, there is a search gap around $\mneutwo - \mlsp \approx m_Z$, where the exclusion reach for $\mneutwo$ is only about $225$ GeV~\cite{CMS:2017sqn}. We find that the search using $\Rmb$ variable with an ISR jet could compete with the tri-lepton search around that region.

This paper is organized as follows. In section~\ref{sec:kinematics}, we discuss the kinematic constraints for the stop pair production in the $W$ corridor with an ISR jet in the semileptonc channel. We define $\Rmb$ for this case and describe how to obtain the minimum and maximum allowed $\Rmb$ values from the constraint equations as the kinematic variables for the stop searches. In section~\ref{sec:Bench}, we investigate the usefulness of the $\Rmbmin$, $\Rmbmax$ variables for the stop searches in the $W$ corridor. A more detailed description of the analyses is given for a chosen benchmark stop mass at 450 GeV and it is shown that  $\Rmbmin$ is quite useful in suppressing certain SM backgrounds. We then perform the study for a series of points along the $W$ corridor to obtain the signal significances in comparison with current search limits.  In section~\ref{sec:chargino} we apply the similar analysis to the chargino (and second neutralino) pair productions and compare with the tri-lepton search limits. Section~\ref{sec:conclusions} contains our conclusions. A detailed description of the solutions for $\Rmbmin$ and $\Rmbmax$ from the kinematic constraints is presented in appendix~\ref{app:technical}. In appendix~\ref{app:validation} we compare the analyses with and without using the $\Rmbmin$ and $\Rmbmax$ variables and show that they indeed can improve the signal significances.

\section{Kinematics and Variables}
\label{sec:kinematics}

For the stop pair production together with and ISR jet, the momentum conservation tells us that 
\begin{equation}
\vec{p}_{T(J_{\text{ISR}})}= - (\vec{p}_{T\tt, 1} +\vec{p}_{T\tt, 2}).
\label{eq:momentum}
\end{equation}
In the $W$ corridor where $\mtt \approx m_W+ m_b+ \mlsp$, the $W$, $b$ and $\lsp$ from the $\tt$ decay will simply be co-moving with the same velocity as their mother particle $\tt$. Therefore we have 
\begin{equation}
\frac{p_{T \lsp, 1(2)}}{p_{T\tt,1(2)}} = \frac{\mlsp}{\mtt} .
\label{eq:mass_ratio}
\end{equation}
Together with Eq.~(\ref{eq:momentum}) we obtain
\begin{equation}
\frac{\mlsp}{\mtt} = - \frac{\vec{p}_{T \lsp,1} + \vec{p}_{T \lsp,2}}{\vec{p}_{T(J_{\text{ISR}})}} \equiv \Rmb ,
\label{eq:Rmb}
\end{equation}
which is the kinematic variable that we would like to use for discriminating stop signals from backgrounds. However, for semileptonic decays, this quantity is not directly measurable, because the neutrino from the $W$ decay also contributes to the missing transverse momentum $\vec{\slashed{p}}_T$ besides the two $\lsp$'s. To obtain $\Rmb$ we need to know the neutrino momentum so that it can be subtracted from $\vec{\slashed{p}}_T$ to get the total $p_T$ of the two $\lsp$'s. 

The neutrino momentum satisfies the two mass shell conditions:
\begin{eqnarray}
p_\nu^2 &= 0 , \label{eq:mass_shell} \\
(p_\ell+p_\nu)^2 &= m_{W}^2.
\label{eq:mass_shell2}
\end{eqnarray}
In addition, the sum of the transverse momenta of the two $\lsp$'s, $\vec{p}_{T\lsp,1}+ \vec{p}_{T\lsp,2}$,  should be antiparallel to the ISR jet. If we decompose $\vec{\slashed{p}}_T$ into components parallel and perpendicular to the $\vec{p}_{T(J_{\text{ISR}})}$ direction, the perpendicular component should be attribute to the neutrino:
\begin{equation}
\slashed{p}_T^\perp = p_{T \nu}^\perp ,
\label{eq:perp}
\end{equation}
which gives us one more constraint on the neutrino momentum once $\slashed{p}_T^\perp$ is determined.
On the other hand, the parallel component receives contributions from both the $\lsp$'s and the neutrino:
\begin{equation}
\slashed{p}_T^\parallel = p_{T \nu}^\parallel + p_{T \lsp, 1}^\parallel + p_{T\lsp,2}^\parallel .
\label{eq:para}
\end{equation}
Using Eq.~(\ref{eq:Rmb}), we can write
\begin{equation}
p_{T\nu}^\parallel = \slashed{p}_{T}^\parallel - (p_{T \lsp, 1}^\parallel + p_{T\lsp,2}^\parallel) = \slashed{p}_{T}^\parallel  + \Rmb\times {p}_{T(J_{\text{ISR}})},
\end{equation}
where the quantities include signs which represent being parallel or antiparallel to the ISR.

We can see that there is one more unknown than the number of kinematic constraint equations, so we can not completely solve the constraint equations to obtain a unique or discrete solution.\footnote{In contrast, in the top corridor there is an additional mass shell condition of the top quark mass, $(p_\ell+p_\nu +p_b)^2 = m_{t}^2$, so the constraint equations can be solved to yield discrete solutions~\cite{Cheng:2016mcw}.} However, the kinematic constraints still limit the solutions to a finite range. As in Ref.~\cite{Cheng:2017dxe}, the minimum and the maximum values of the allowed range of $\Rmb$, $\Rmbmin$ and $\Rmbmax$, provide potential variables for the signal and background discrimination. The detailed computation of $\Rmbmin$ and $\Rmbmax$ is presented in Appendix~\ref{app:technical}. The combination of the kinematic constraints gives a quadratic equation for one neutrino component. To have real solutions, the discriminant of the quadratic equation is required to be $\geqslant 0$ . The discriminant is also quadratic in $\Rmb$, so the two solutions of discriminant = 0 give $\Rmbmin$ and $\Rmbmax$.

In the above discussion, the information of the $b$-jets from the stop decays is not used at all. In the $W$ corridor, the $b$-jets are typically too soft to be identified or to be useful. As a result, the same analysis also applies to the chargino pair production with the chargino decays to $W + \lsp$ and $\mchar \approx m_W+\mlsp$. In SUSY, there is usually at least one neutralino ($\tilde{\chi}_2^0$) having a similar mass as that of the chargino, so the neutralino-chargino pair production is also important at the same time. If $\tilde{\chi}_2^0$ decays to $Z^{(\ast )}(h^{(\ast)})+ \lsp$ with $Z^{(\ast )}(h^{(\ast)})$ decays hadronically and the $W$ from the chargino decays leptonically, our analysis can also apply. For the chargino-neutralino production, the trilepton search traditionally provides the strongest constraint. We will perform a study based on our method in Sec.~\ref{sec:chargino} to compare it with the current trilepton search bound.

\section{Stop Searches along the W Corridor}
\label{sec:Bench}

\subsection{Signal and Background Generations}
We use MadGraph 5~\cite{Alwall:2014hca} and Pythia 6~\cite{Sjostrand:2006za} to generate both the background and the signal events. MLM matching scheme~\cite{Mangano:2002ea} is applied for both the SM background and the SUSY signal production in order to prevent double-counting between the matrix elements and the parton shower. The detector simulation is performed by Delphes 3~\cite{deFavereau:2013fsa}, using the anti-$k_t$ jet algorithm~\cite{Cacciari:2008gp} with the parameter $R=0.5$. For the signals, the production cross sections are normalized to 13 TeV NLO+NLL results~\cite{Borschensky:2014cia}. The $b$-jet tagging efficiency is taken to be the same as one of the benchmark operating points shown in \cite{btagging}, with the maximum efficiency $\approx 77\%$.

Since we require exactly one lepton, large MET and hard extra jets, a number of SM processes can be responsible for such a final state. According to similar/related collider studies~\cite{Sirunyan:2017mrs,Aaboud:2017bac}, we expect SM $t\bar{t}$, $tW$, $W+$ jets and di-boson events to be our main backgrounds, since they can naturally provide a lepton and MET with large production rates. 
  Other backgrounds, such as $tt+W/Z$, $Z+$ jets and $tb$ events either suffer from low production cross sections or low signal efficiencies. All SM backgrounds mentioned above are generated by the method aforementioned. Besides the SM backgrounds, the dileptonic decay of $\tilde{t}\tilde{t}^\ast$ can be an irreducible background to the signal. However, this process has a much smaller cross section compared to the SM backgrounds and can be ignored for the rest of our discussion.

\subsection{Event Selection}

For our benchmark studies, all events must satisfy the preliminary selection as described below.  Each event is required to have at least 2 jets, 0 tau-tagged jet and exactly 1 isolated lepton with 20 GeV $< p_T <100$~GeV and $|\eta|<2.5$. The upper limit of the transverse momentum is imposed because our signal comes from a compressed spectrum and the $W$ bosons are not very boosted. Since we need a hard ISR jet, the leading jet is required to satisfy $p^{j_1}_{T}>150$~GeV while the rest of the jets must have $p^{j_2,j_3...}_{T}\leq 150$~GeV.  A jet must have $p_T>20$~GeV and $|\eta|<5$ to be considered in the later analysis. The signal event is expected to have a substantial amount of missing transverse momentum from the neutralinos recoiled against the ISR jet. On the other hand, the missing transverse momentum of the SM backgrounds mainly comes from neutrinos. The requirement $\slashed{p}_T>200$~GeV is applied to eliminate most of  the backgrounds. For top-related backgrounds, $t\to W+b$ decay provides hard $b$ jet(s) in the final states, thus we veto all tagged $b$ jets to reduce the impact of these backgrounds. Furthermore, to control the large $W+$ jets background where the missing transverse energy is due to a single neutrino from the $W$ decay, a cut on the transverse mass $M_T \ge 100$~GeV is imposed. It proved to be very effective to suppress $W$+ jets and other semileptonic backgrounds. 

In our signal events, the neutralinos are recoiled against the ISR.  To make sure that the leading jet is antiparallel to the sum of neutralinos' momenta and is not from the decay of the stops, we require that $|\phi_{j_1}-\phi_{\rm MET}|\ge 2$ and $\Delta R_{j_1,\ell} \ge 1.5$. To take into account the cases where there is more than one ISR jet, we define $p_{T(J_{\text{ISR}})}$ to be the vector sum of all jets' $p_T$ that are inside the $\Delta R \le 2$ cone with the leading jet $j_1$ and outside $\Delta R \ge 1$ with the lepton.

\subsection{Stop Benchmark Study}
\label{sec:Stop450}

For an illustration, we describe in detail the analysis for a benchmark point ``Stop-450,'' $\mtt=450$~GeV, $\mlsp =363$~GeV along the $W$ corridor in the parameter space. For simplicity, we assume all other supersymmetric particles are decoupled and the stop's decay branching ratio to $b+W+\tilde{\chi}$ through an off-shell top is $100\%$. Since the spectrum of interest is very compressed, the searches based on the $M_{T2}$ types of variables are ineffective, which motivates us to explore the usefulness of the $\bar{R}_M$ type of variables. 

\begin{figure}
\captionsetup{singlelinecheck = false, format= hang, justification=raggedright, font=footnotesize, labelsep=space}
\includegraphics[scale=0.5]{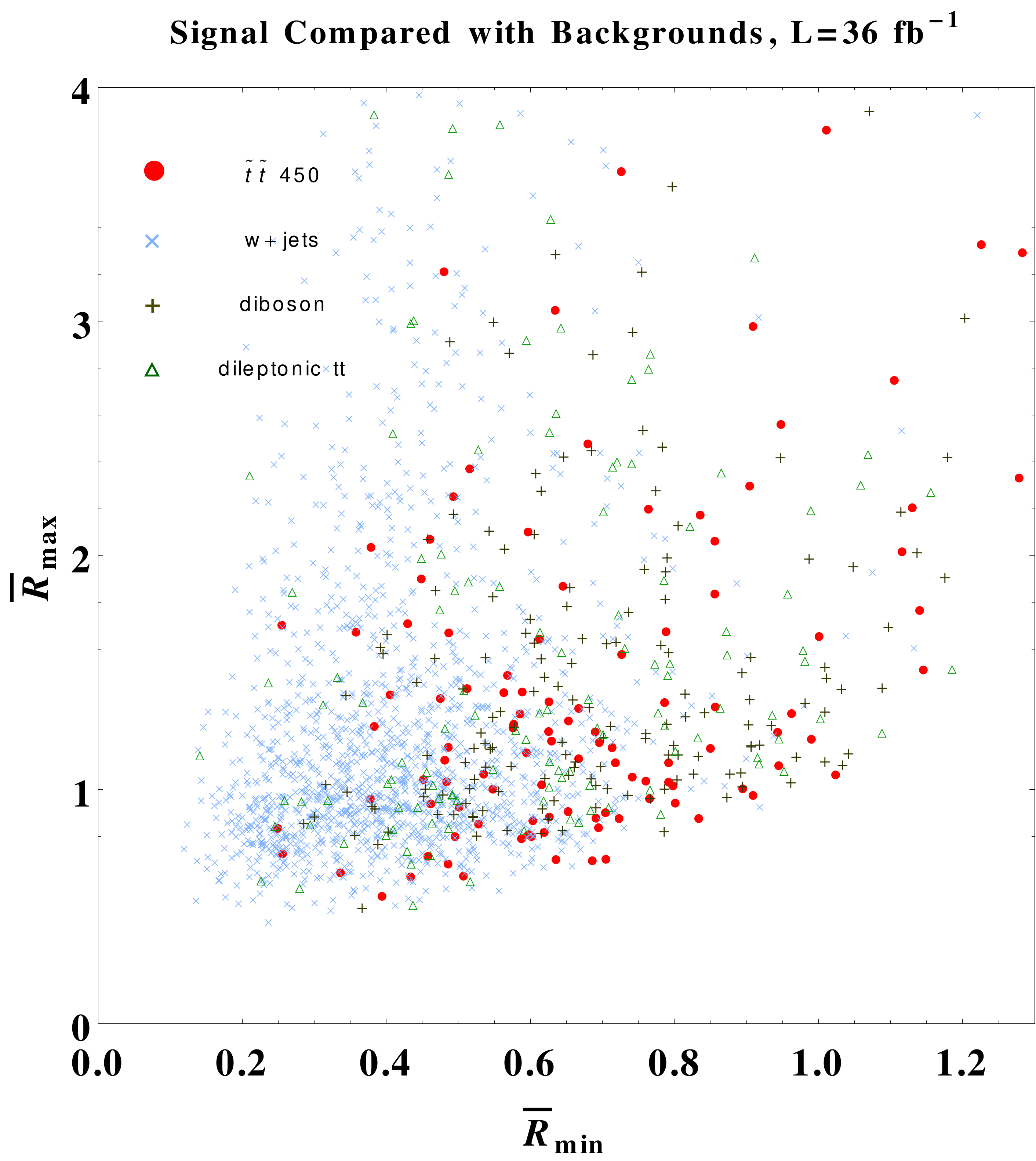}
\caption{The Stop-450 benchmark signal and background distributions on the $\Rmbmin - \Rmbmax$ plane after the preliminary selection. Only events with real solutions are shown. $W+$jets is the dominant background at this stage. Diboson and dileptonic $t\bar{t}$ backgrounds tend to have similar distribution as the signal, therefore are more difficult to remove. They remain as significant components of the background at the end.}
\label{fig:first}
\end{figure}

In Fig.~\ref{fig:first} we plot the two-dimensional $\Rmbmin - \Rmbmax$ distributions for the signal benchmark point Stop-450 and SM backgrounds after the preliminary selection. As expected, the stop signal events mostly distribute at larger $\bar{R}_M$ values, while SM backgrounds appear at both small and large $\bar{R}_M$ values.

A closer look at the data shows that backgrounds with one single neurtrino as the source of their missing energy tend to give small $\bar{R}_M$.  These ``mono-neutrino'' backgrounds include $W$ + jets and semileptonic $t\bar{t}$ production, where there is no other invisible particle except one neutrino. In principle, these backgrounds should allow $\Rmb=0$ as a solution from the equation 
\begin{equation}
-\Rmb \times {p}_{T(J_{\text{ISR}})}+p_{T_\nu}=\slashed{p}_T.
\label{eqn:eq1}
\end{equation} 
However, if the measured $\slashed{p}_T$ purely comes from $p_{T_\nu}$, the corresponding transverse mass $M_T$ is bounded by the $W$ mass,
\begin{equation}
M_T(\ell,\text{MET})=\sqrt{2p_{T \ell}\slashed{p}_T(1-\cos\Delta\phi_{\ell,\text{MET}})}\leqslant(p_\ell+p_\nu)^2=m^2_W,
\label{eqn:eq2}
\end{equation}
which would have been removed by the $M_T \ge 100$~GeV cut. The events that passed the cut must have some additional $\slashed{p}_T$ due to mismeasurements or lost particles, which generally renders a positive $\Rmb$ as shown in the figure. We also see that $\Rmbmin$ is a more useful variable to suppress these backgrounds than $\Rmbmax$ which has a wider distribution.

On the other hand, those backgrounds which can produce more than one neutrino such as $WZ$ and dileptonic $t\bar{t}$ likely give larger $\bar{R}_M$. This is because when leptons are not identified by the detector or neutrinos are pair produced from $Z$, the extra neutrino or lost lepton momentum becomes part of the MET. Consequently, the assumptions mentioned in Sec.\ref{sec:kinematics} are violated. More often than not, these extra neutrinos are produced in the different hemisphere of the ISR system, the MET then becomes larger and $\bar{R}_M$ tends to be more positive. Such an effect due to extra neutrinos also occurred for the top-corridor stop search~\cite{Cheng:2016mcw}. As a result, this type of background overlaps more with signals and the $\Rmb$ variables are less effective in removing them. Fortunately they are subdominant compared to the $W+$jets background which can be effectively suppressed by the $\Rmb$ variables.

\begin{figure}
\captionsetup{singlelinecheck = false, format= hang, justification=raggedright, font=footnotesize, labelsep=space}
\includegraphics[scale=0.5]{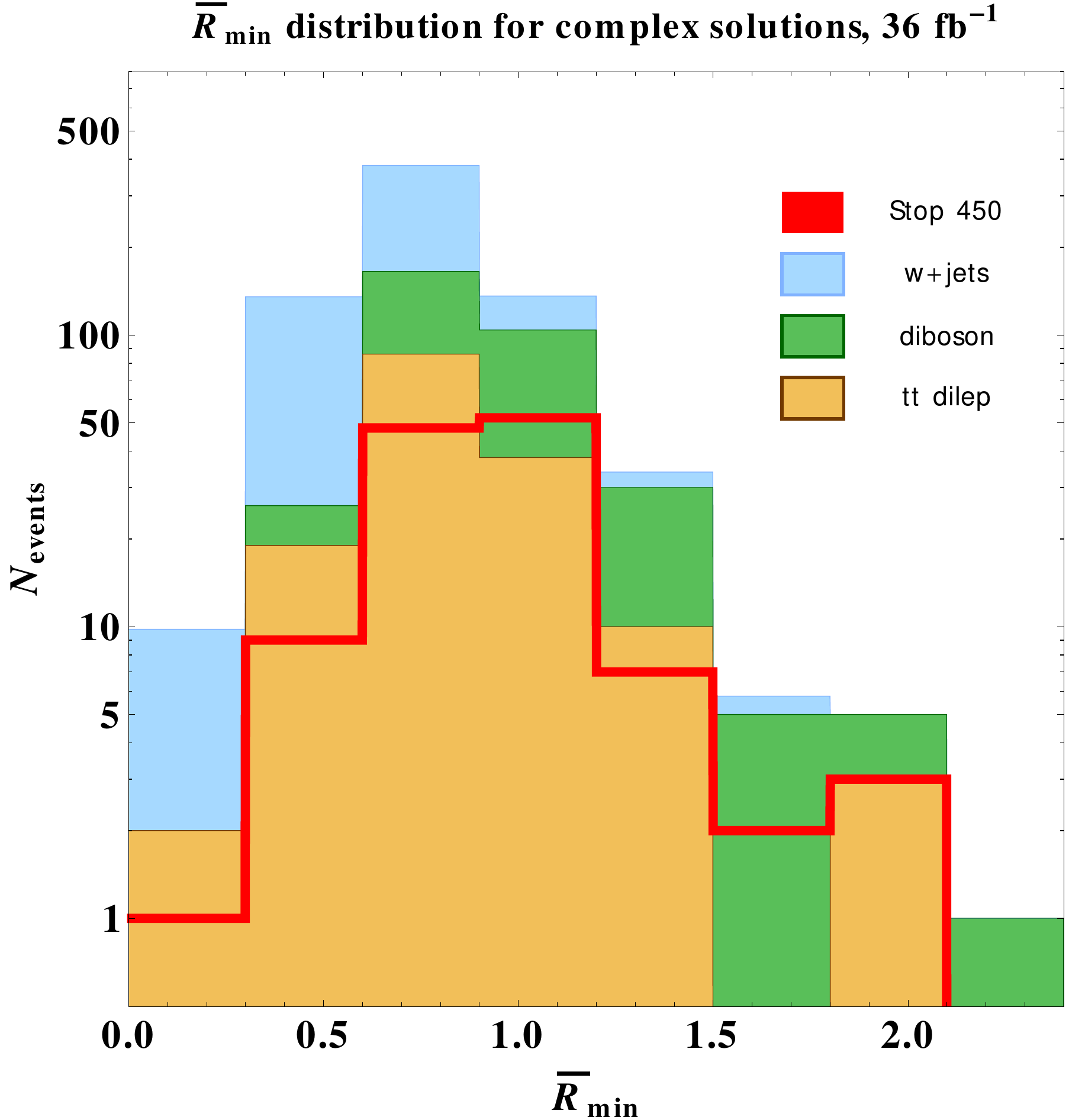}
\caption{The Stop-450 benchmark signal and background distribution for the complex $\bar{R}_M$ solutions.}
\label{fig:stop-complex}
\end{figure}

For events that do not yield real $\Rmb$ solutions, one possibility is to simply discard them. However, there are quite some signal events which do not have real solutions, which may be caused by mismeasurements. In those cases, one might hope that the real part of the complex solutions of $\Rmb$ gives a reasonable approximation to the true $\Rmb$ value. To check if this helps to increase the signal significance, for those events that give complex solutions for $\Rmb$, we simply define $\Rmbmin (=\Rmb)$ to be the real part of the complex solutions and plot its distributions for the signal and backgrounds in Fig.~\ref{fig:stop-complex}. The empirical variable turns out to be also useful as one can see that the background events (especially the dominant $W+$jets) have a lower distribution than that of the signal, although it is not as good as $\Rmbmin$ in the real solution case.

\begin{figure}[t]
\captionsetup{singlelinecheck = false, format= hang, justification=raggedright, font=footnotesize, labelsep=space}
\includegraphics[scale=0.35]{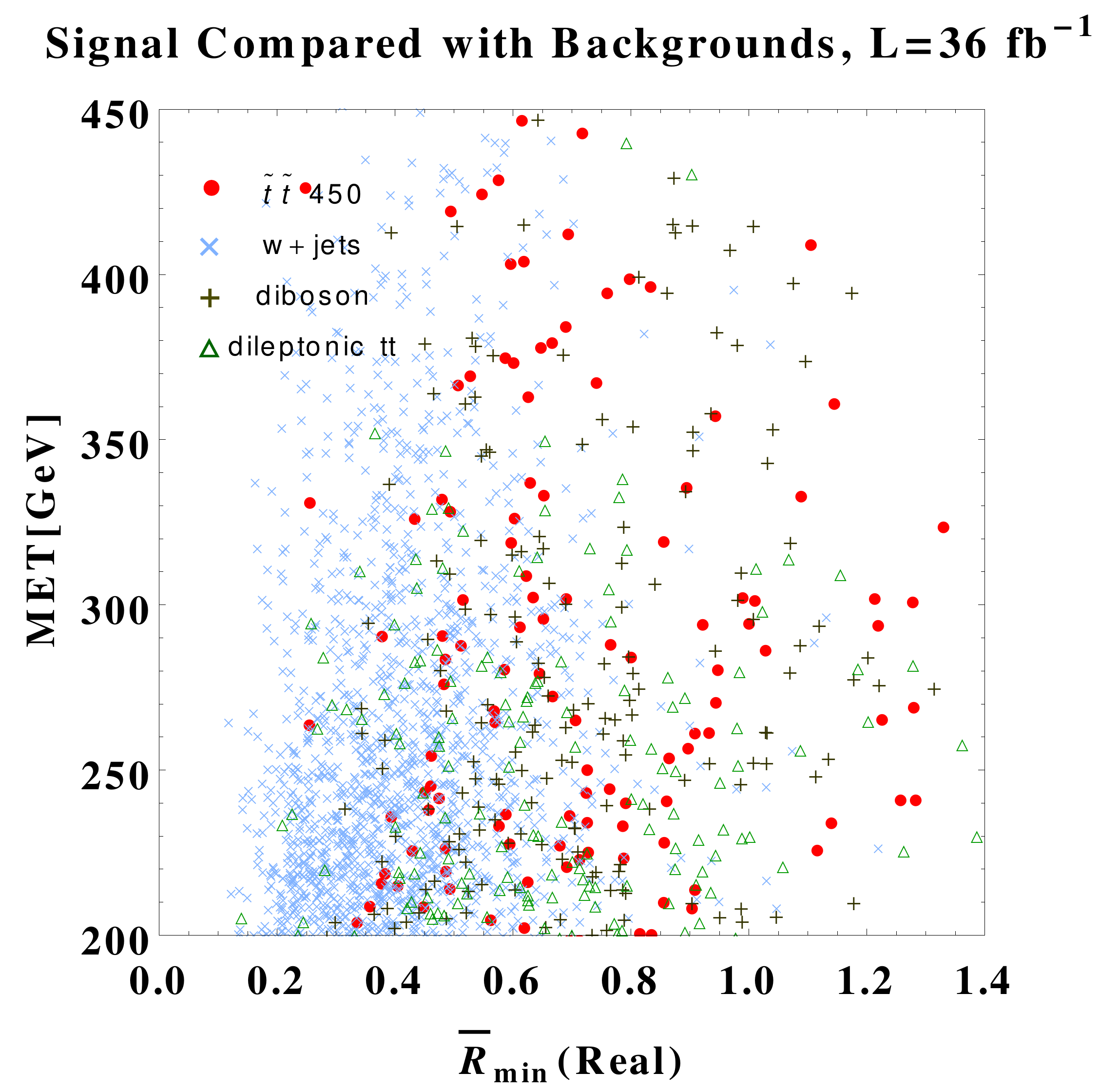}
\includegraphics[scale=0.35]{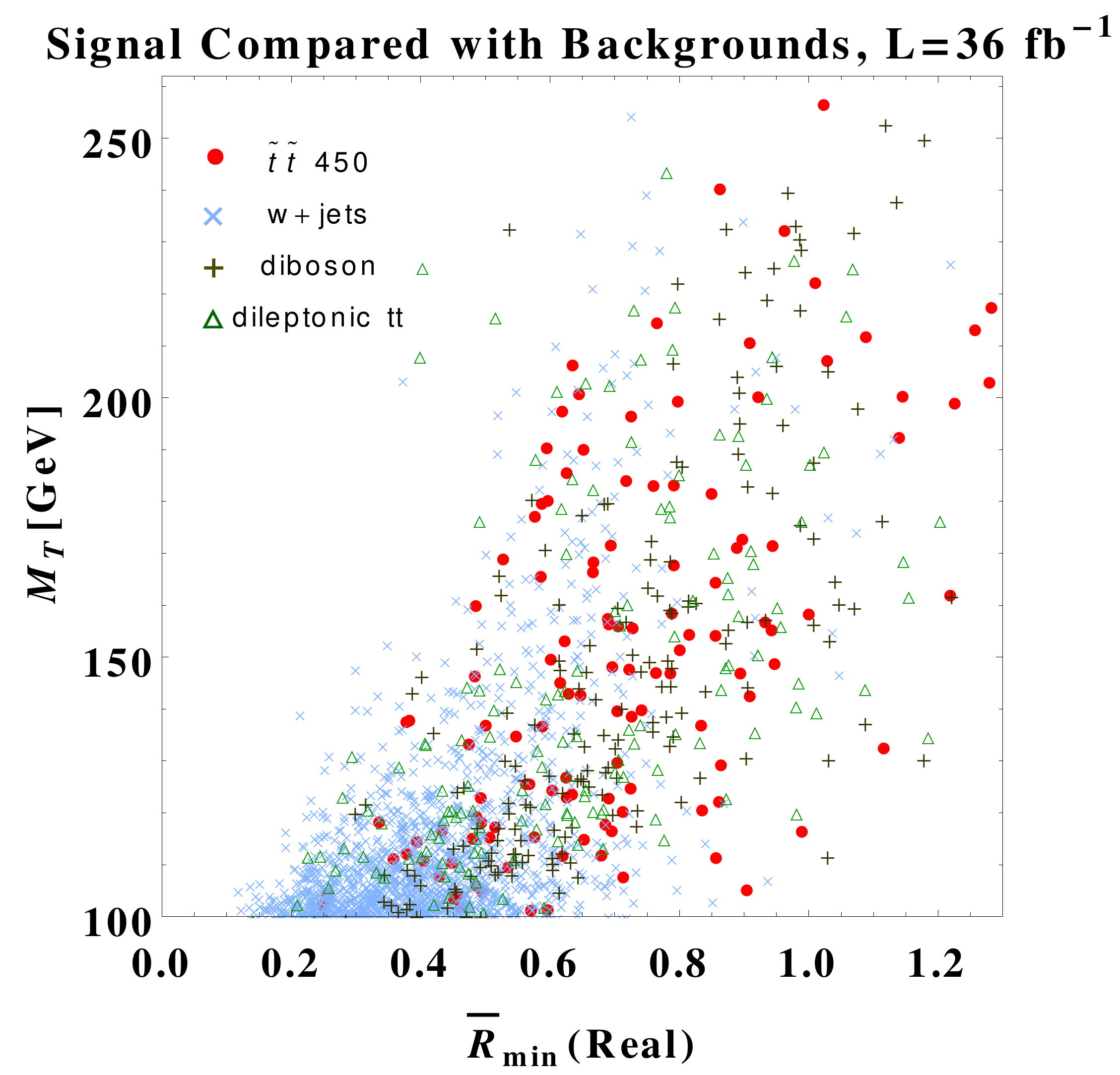}
\caption{MET vs.\ $\Rmbmin$ and $M_T$ vs. $\Rmbmin$ distributions for events with real solutions.}
\label{fig:seven-real}
\end{figure}

\begin{figure}[t]
\captionsetup{singlelinecheck = false, format= hang, justification=raggedright, font=footnotesize, labelsep=space}
\includegraphics[scale=0.35]{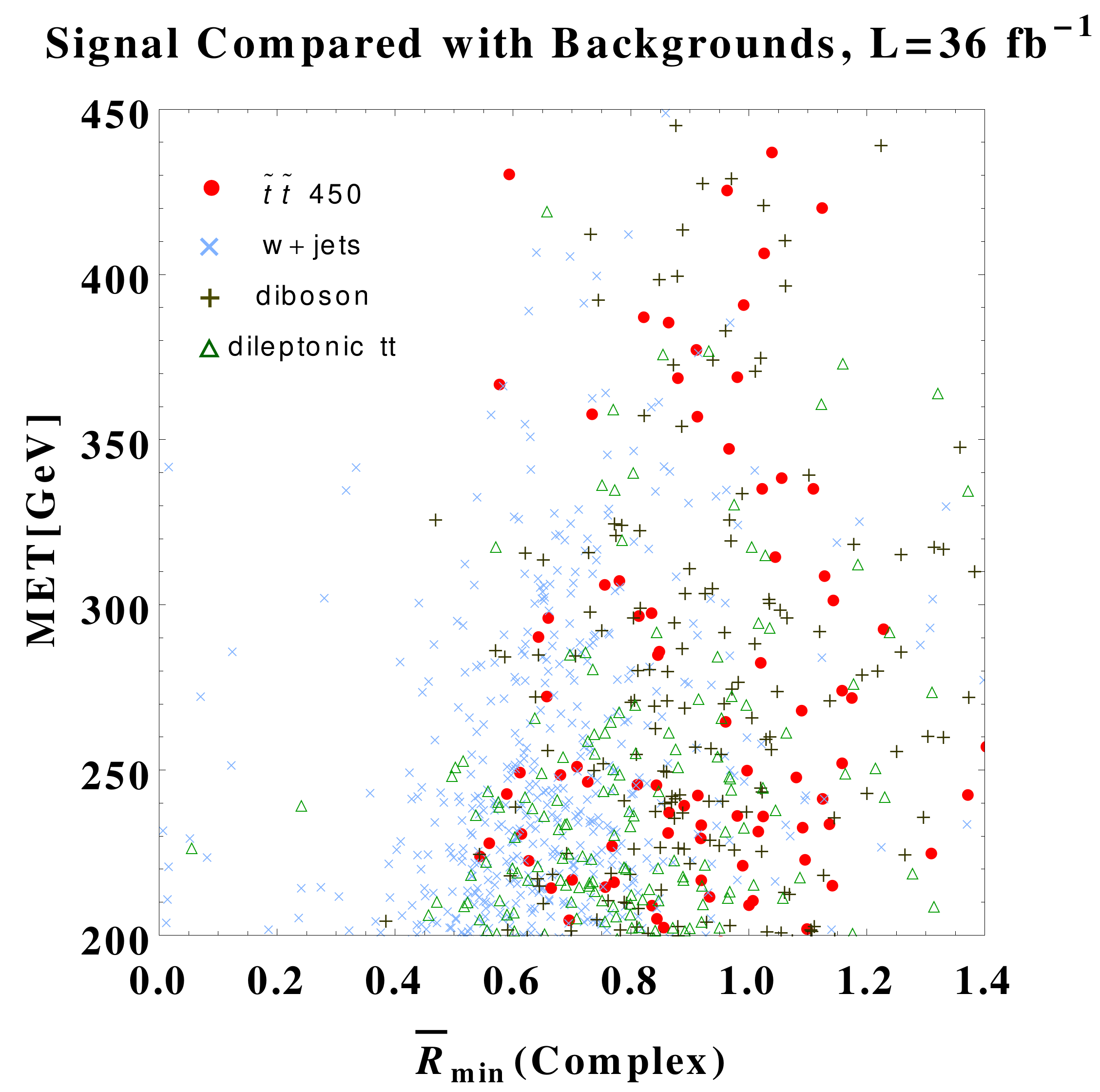}
\includegraphics[scale=0.35]{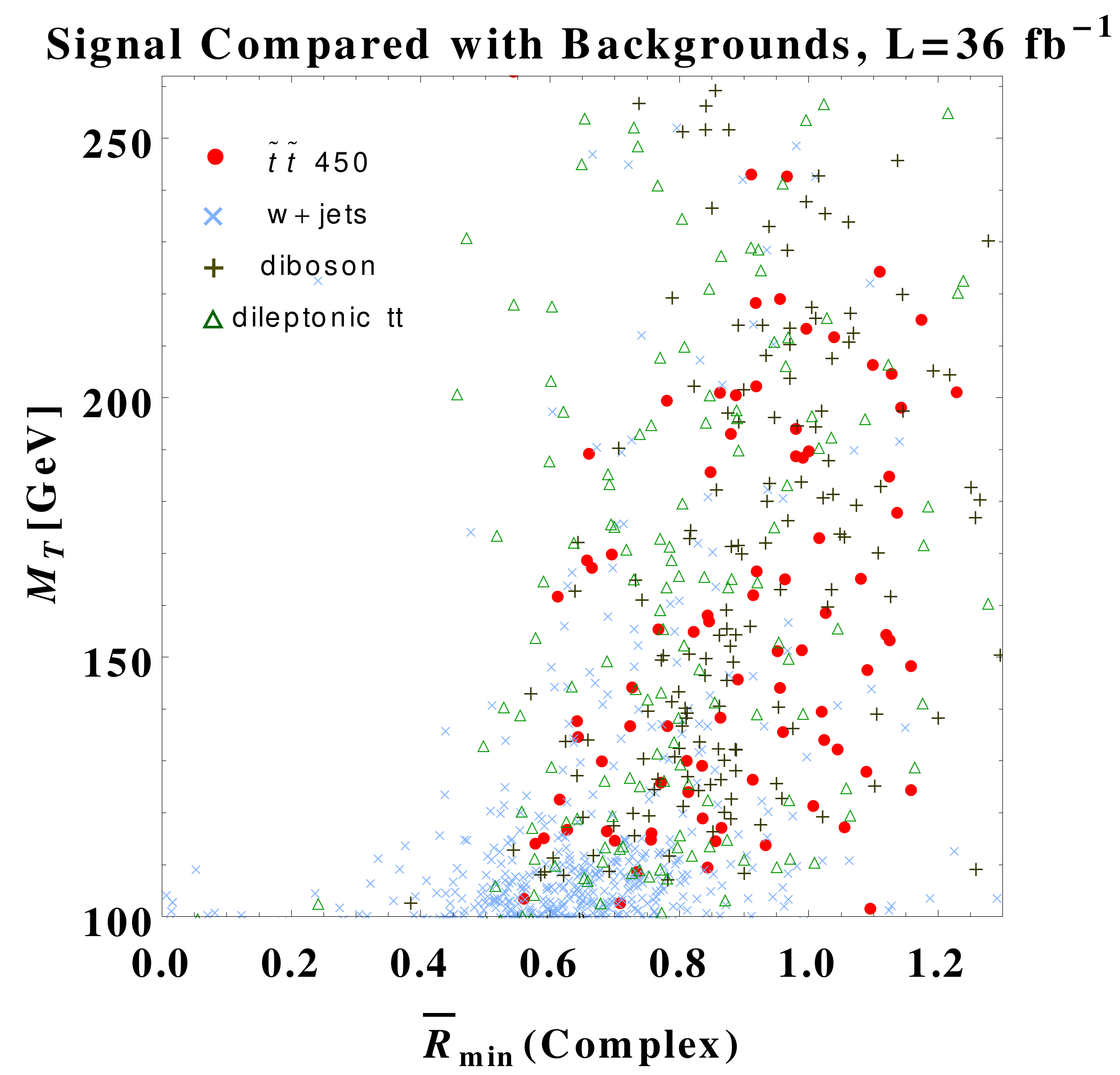}
\caption{MET vs.\ $\Rmbmin$ and $M_T$ vs. $\Rmbmin$ distributions for events with complex solutions.}
\label{fig:seven-complex}
\end{figure}

After seeing the usefulness of the $\Rmbmin$ variable, we plot the signal and background distributions against other variables, MET and $M_T$ in Figs.~\ref{fig:seven-real} and \ref{fig:seven-complex}. We see that the dominant backgrounds generally also have lower values in MET and $M_T$. 
Based on these distributions, we make the following cuts to select two signal regions:
\begin{itemize}
\item for events with 200 GeV $\leqslant$ MET $<$ 300~GeV, $\Rmbmin \geqslant 0.7$ and $M_T>150$~GeV are required (SRL);
\item for events with MET $\geqslant 300$~GeV, $\Rmbmin \geqslant0.5$ and $M_T>120$~GeV are required (SRH).
\end{itemize}
For smaller MET where the backgrounds are large, we impose harder cuts on $\Rmbmin$ and $M_T$ to reduce the background events. For large MET, $\Rmbmin$ and $M_T$ cuts can be relaxed a bit to allow more signal events to pass them.

The numbers of events passing the above cuts for the benchmark signal and SM backgrounds, normalized to an integrated luminosity of 36~fb$^{-1}$ are shown in Table~\ref{tab:1}.
\begin{table}[ht]
\captionsetup{singlelinecheck = false, format= hang, justification=raggedright, font=footnotesize, labelsep=space}
\begin{small}
\begin{tabular}{l|cccccc}
         & Initial & Prelim & SRL(R)  & SRH(R) & SRL(C) & SRH(C) \\ 
\hline\hline
$\tilde{t}\tilde{t}^\ast$ 450~GeV & $1.11\times 10^4$ & $2.98\times 10^2$ & 20.6 & 51.6 & 25.4 & 38.0 \\
\hline
\hline
W+jets  & $ 1.94\times 10^6$ & $1.61\times 10^3$ & 25.7 & 42.8 & 19.4 & 23.3 \\
di-boson & $4.37 \times 10^5$ &  $3.65 \times 10^2$ & 33.1 & 56.0 & 61.1 & 56.2 \\
$t\bar{t}$ (Dilep) & 1.97 $\times 10^6$ & $3.46 \times 10^2$ & 40.5 & 18.6 & 62.3 & 19.6 \\
$t\bar{t}$ (Semilep) & 7.86 $\times 10^6$ & 9.2 & 0 & 0 & 0 & 0 \\
SM other & 3.50 $\times 10^6$ & $9.23 \times 10^1$ & 9.8 & 7.5 & 15.3 & 3.1\\
\hline\hline
SM total & $1.58\times 10^7$ & $24.5\times 10^3$ &  109 & 125 & 158 & 102  \\
\end{tabular} 
\end{small}
\caption{The cut flow for the stop signal and backgrounds, assuming an integrated luminosity of 36 fb$^{-1}$. (R) and (C) indicate events with real and complex solutions respectively.}
\label{tab:1}
\end{table}
After the final selection, the largest background comes from the diboson which is dominated by $WZ$+jets. This is because $Z$ can decay to two neutrinos which imitates the neutralinos of the signal events.

To calculate the signal significances for the benchmark models, we use the likelihood method with the assumption
that the overall number of background events in each signal region respects the normal distribution with a fractional uncertainty $\sigma_{B}\propto B$. The likelihood is defined to be 
\begin{equation}\label{likelihood}
Q=\frac{\int \mathcal{L}(S+B,S+B')P(B')dB'}{\int \mathcal{L}(S+B,B')P(B')dB'},
\end{equation}
where $S$ and $B$ are corresponding numbers of signal and background events, $\mathcal{L}(x,\mu)=\frac{\mu^{x}e^{-\mu}}{x!}$, and $P(B)$ is the normalized normal distribution with the mean $B$ and a standard deviation $\sigma_B$. The final significance from this method is simply given by $\sqrt{2\log(Q)}$. 
For the case with no systematic error, $\sigma_B=0$, this equation simply reduces to the standard formula~\cite{Cowan:2010js}:
\begin{equation}
\sigma = \sqrt{2\left[(S+B)\log\left( \frac{S+B}{B} \right)-S\right]}.
\label{significance}
\end{equation}
For the Stop-450 benchmark, we get a significance of $4.3\,\sigma\,(6.3\,\sigma)$  for 36~fb$^{-1}$ with (without) a 10$\%$ background uncertainty. 
For current LHC SUSY searches also using one-lepton final states~\cite{Aaboud:2017bac}, the systematic uncertainties for different signal regions vary from $\sim 10\%$ to $\sim 30\%$, which  mainly comes from uncertainties in modeling the SM backgrounds and MC simulations rather than the experimental uncertainties. In the future when the integrated luminosity increases from $36~\ifb$ to $300~\ifb$, we expect that the systematic uncertainties will further decrease but the actual numbers are hard to predict. Here we use a 10\% background systematic uncertainty to demonstrate its impact on the signal significances. The results with different background uncertainties can also be obtained easily from the numbers in Table~\ref{tab:1}.

One question of the analysis is how much the $\Rmb$ variables help the stop search in this case. One can imagine that a variable defined by the ratio MET/ISR gives a simple approximation of $\Rmb$ and hence the search can be done with the standard simple variables MET, ISR, and $M_T$. To check this we plot the MET vs.\ ISR distributions of the signal and background in Fig.~\ref{fig:second}.  One can see that there is some separation between signal and backgrounds, but compared to Fig.~\ref{fig:first} it does not seem to be as good. In Appendix~\ref{app:validation}, we perform an analysis without using the $\Rmb$ variables and find that indeed the signal significance is substantially inferior to the result obtained here with the $\Rmb$ variables. 
\begin{figure}[t]
\captionsetup{singlelinecheck = false, format= hang, justification=raggedright, font=footnotesize, labelsep=space}
\begin{subfigure}[b]{0.45\textwidth}
\includegraphics[width=8.5cm]{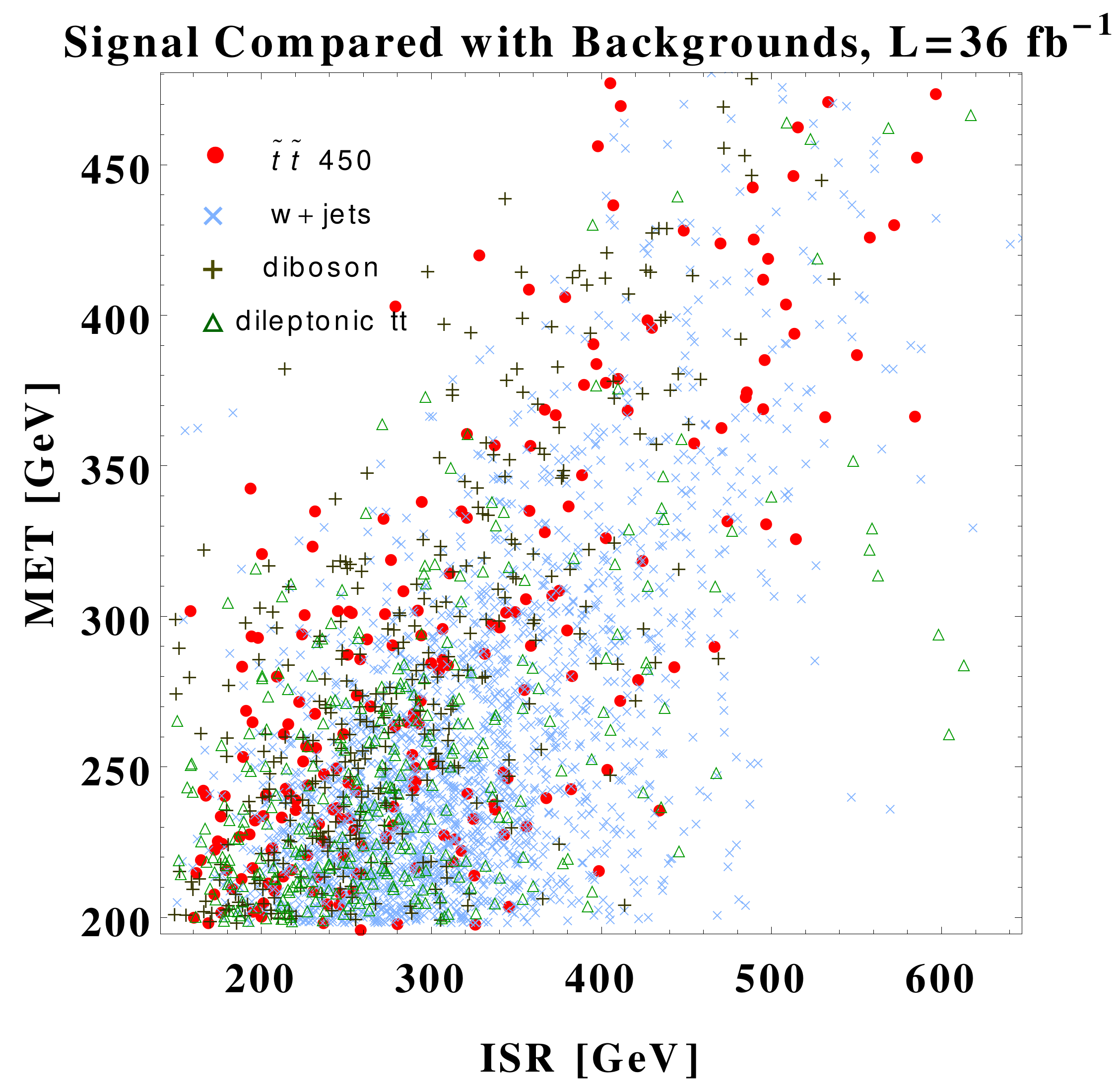}
\end{subfigure}
\caption{MET vs.\ ISR distributions for Stop-450 signal and backgrounds after the preliminary selection.}
\label{fig:second}
\end{figure}

\subsection{Stop Results at LHC 13TeV}
\label{sec:results}

\begin{figure}
\captionsetup{singlelinecheck = false, format= hang, justification=raggedright, font=footnotesize, labelsep=space}
\includegraphics[scale=0.6]{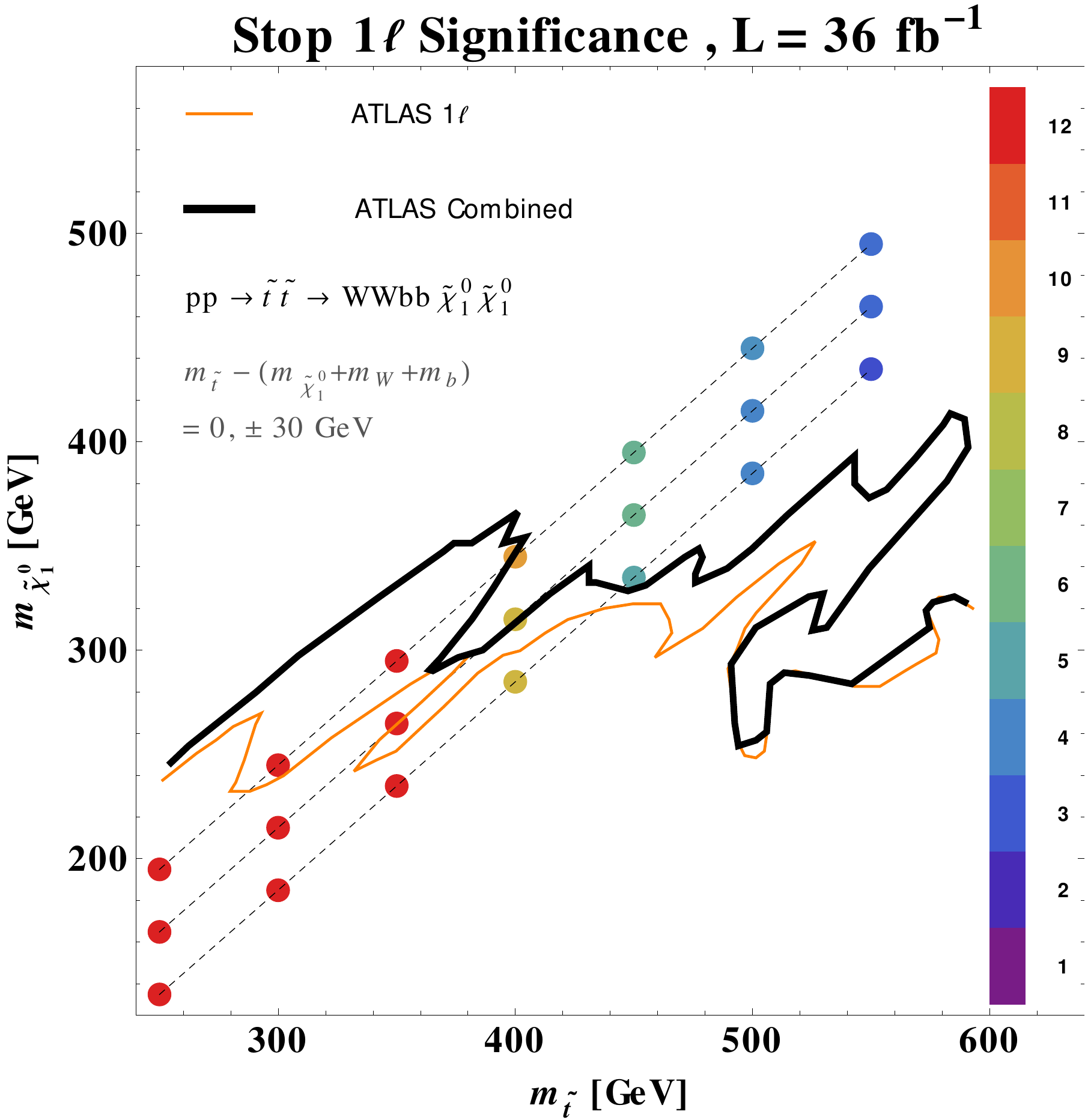}
\caption{The signal significances based on the analyses in this study for points along the $W$ corridor in the parameter space compared to the current experimental exclusion limit at LHC13.}
\label{fig:fourth}
\end{figure}
The discussion of the last subsection demonstrated that the analysis based on the $\Rmb$ variables with a hard ISR can yield a large signal significance for a 450~GeV stop in the $W$ corridor with 36 $\ifb$ integrated luminosity. To study the reach of this method, we perform the same analysis for a series of points along the $W$ corridor. The optimal signal regions may depend on the mass points, but for simplicity and easy comparison we use the same signal regions defined in the previous subsection. The results are shown in Table~\ref{tab:2} and Fig.~\ref{fig:fourth}. 
We find that a stop mass below $550$~GeV in the $W$ corridor with the assumed decay mode can be excluded at the 95\% CL by this method.
In comparison, the current ATLAS 1-lepton analysis only excludes stop mass up to $\sim 340$~GeV in the $W$ corridor~\cite{Aaboud:2017bac}, The ATLAS 2-lepton analysis can exclude stop mass up to $\sim 430$~GeV just above the sum $m_W+m_b+\mlsp$, but leaves a gap below that where the reach degraded to $\sim 360$~GeV~\cite{Aaboud:2017nfd}. They are far beneath the potential reach of the new approach studied here.
\begin{small}
\begin{table}[ht]
\captionsetup{singlelinecheck = false, format= hang, justification=raggedright, font=footnotesize, labelsep=space}
\begin{tabular}{c|ccccccc}

       $\mtt$ (GeV)  & 250 & 300 & 350 & 400 & 450 & 500 & 550 \\ 
\hline
\hline
$\sigma_{(\mtt-\mlsp-m_W-m_b=0)}$ & 27.9(37.4) & 18.0(25.4) & 11.8(17.2) & 7.0(10.2) & 4.3(6.3) & 2.7(3.9) & 2.1(3.0) \\

$\sigma_{(\mtt-\mlsp-m_W-m_b=-30)}$ & 37.6(47.8)  & 20.5(27.9) & 10.9(15.8) & 7.9(11.5) & 4.2(6.1) & 3.0(4.3) & 2.2(3.2)  \\

$\sigma_{(\mtt-\mlsp-m_W-m_b=30)}$  & 20.1(28.0)  & 14.2(20.7) & 10.8(16.1) & 7.0(10.2) & 3.7(5.4) & 2.7(3.9) & 1.6(2.3)  \\
\end{tabular} 
\caption{Stop search significances with (without) a 10\% systematic background uncertainty using $\Rmb$ variables in the $W$ corridor region, assuming 36 $\ifb$ integrated luminosity at LHC 13 TeV }
\label{tab:2}
\end{table}
\end{small}

We are also interested in the mass parameter region slightly away from the $\mtt - \mlsp = m_W+m_b$ line to see the coverage of our method. We performed the same analysis for points along the lines of $\mtt - \mlsp = m_W+m_b \pm 30$~GeV. The results are also shown in Table~\ref{tab:2}. Away from the $\mtt - \mlsp = m_W+m_b$ line, some of the kinematic assumptions used in Sec.~\ref{sec:kinematics} are no longer valid. For instance, when the mass gap between $\tt$ and $\lsp$ is larger than $m_W+m_b$, the $W$ bosons are still on-shell, so Eq.~(\ref{eq:mass_shell2}) still holds. However, the neutralinos would no longer be static in the rest frame of the stops and consequently the sum of their momentum may no longer be strictly antiparallel to the ISR. Thus, our assumption that the neutrino is solely responsible for $\slashed{p}^\perp_T$ is no longer justified. This could further smear the $\Rmb$ distribution for the signal, hence reducing its discriminating power. We see that the significances for the points on the $\mtt - \mlsp = m_W+m_b + 30$~GeV line are generally somewhat worse than those on the $\mtt - \mlsp = m_W+m_b$ line for the same stop mass. On the other hand, for a stop lighter than $\mlsp+m_W+m_b$, the stop goes through the 4-body decay and the mass shell condition Eq.~(\ref{eq:mass_shell2}) is no longer valid. However, the mass ratio ${\mlsp}/{\mtt}$ is larger for the same $\mtt$. The distribution of $\Rmb$ for the signal also shifts to larger values, resulting in better separation from the backgrounds. The signal significances along the $\mtt - \mlsp = m_W+m_b - 30$~GeV line are still comparable to the points along the $\mtt - \mlsp = m_W+m_b$ line. From these results, we conclude that this new approach can apply to a quite wide region around the $W$ corridor and will extend the coverage on the search gap present in the current experimental analyses.

\section{Chargino and Neutralino Searches along the W Corridor}
\label{sec:chargino}
\begin{figure}[h]
\captionsetup{singlelinecheck = false, format= hang, justification=raggedright, font=footnotesize, labelsep=space}
\includegraphics[scale=0.45]{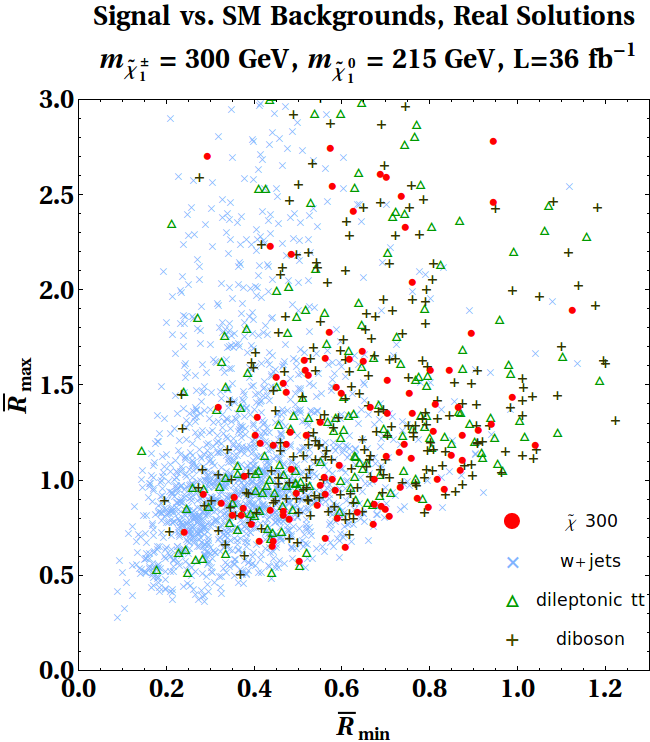}
\caption{300~GeV chargino/neutralino signal vs. various SM backgrounds after preliminary selection rules are applied, only events with real solutions are shown.  $W$+ jets seem to be dominant at this stage, and will remain important after further cuts. Moreover, diboson and dileptonic $tt$ backgrounds tend to have similar distribution as the signal, therefore are significant at the end.}
\label{fig:charginoR}
\end{figure}
The compressed chargino/neutralino that decay via $\tilde{\chi}_1^{\pm} \to W^{\pm}+\lsp$ or $\tilde{\chi}_2^0 \to Z+\lsp$ can also give the $\ell$ + jets + MET final state without $b$ jets, thus similar analysis can also be applied to chargino searches along the $W$ corridor. Due to the smaller production cross section, the experimental exclusion limit on $\mchar$ and $\mneutwo$ is weaker compared to stop searches. 
The current (36 $\ifb$) reach of $\tilde{\chi}^{\pm}_1$ and $\nlsp$ is less than $250$~GeV in the compressed region from CMS in the $3\ell$ channel, assuming $\tilde{\chi}^{\pm}_1$ and $\nlsp$ are wino-like and decay to $W$ and $Z$ plus $\lsp$ respectively~\cite{Sirunyan:2017lae,CMS:2017sqn}. (The higgsino-like $\tilde{\chi}^{\pm}_1$, $\nlsp$ can also be constrained as long as they have the same spectrum and decay final states, but their production rate will be a few times smaller.) There is a gap around $\mneutwo - \mlsp \approx m_Z$ where the limit further degrades to $\sim 225$~GeV. The limit from the current ATLAS analysis is even weaker~\cite{ATLAS:2017uun}. There is no chargino/neutralino search using the $1\ell$ channel from either ATLAS or CMS in the compressed region, which motivates us to explore the usefulness of the approach using $\bar{R}_M$ in the chargino search.

We start with a benchmark (C1N2-300) of 300 GeV degenerate $\tilde{\chi}_1^{\pm}$, $\tilde{\chi}_2^0$, with the production rate taken to be wino-like. The LSP $\lsp$ is assumed to be bino-like and has a mass 215 GeV.  $\tilde{\chi}_1^{\pm}$, $\tilde{\chi}_2^0$ are produced through the electroweak process, then decay to $W^{\pm} (Z)+\lsp$ with 100\% branching ratio. The preliminary selection rules are mostly the same as ones applied in the stop study of the previous section, except for the MET requirement. Because the reach in the mass spectrum will be weaker than the stop case due to the smaller production cross section, for the same ISR momentum the recoiled momentum carried by the two $\lsp$'s will be lower for smaller $\mlsp$. With that observation, we lower the MET requirement to $\geqslant 180$~GeV. 

In Fig.~\ref{fig:charginoR} we plot the two-dimensional $\Rmbmax$  vs. $\Rmbmin$ distributions of 300~GeV $\tilde{\chi}_1^{\pm} - \tilde{\chi}_2^0$ 1 lepton signal vs.\ various SM backgrounds after the preliminary selection, normalized to 36~$\ifb$.  Compared to Fig.~\ref{fig:first}, we can see that signal events of C1N2-300 generally have smaller $\Rmb$ range relative to those of the Stop-450 signal, in accordance with our expectation. Specifically, the theoretical $\Rmb$ value for C1N2-300 is $\Rmb^{\rm theory}={\mlsp}/{\mchar}\approx0.62$, smaller than that of Stop-450 where $\Rmb^{\rm theory}={\mlsp}/{\mtt}\approx0.81$. Similar to the stop case, the more useful variable is $\Rmbmin$ as it separates the signal and backgrounds better than $\Rmbmax$, as we can see from~Fig.~\ref{fig:charginoR}.

\begin{figure}
\captionsetup{singlelinecheck = false, format= hang, justification=raggedright, font=footnotesize, labelsep=space}
\includegraphics[scale=0.35]{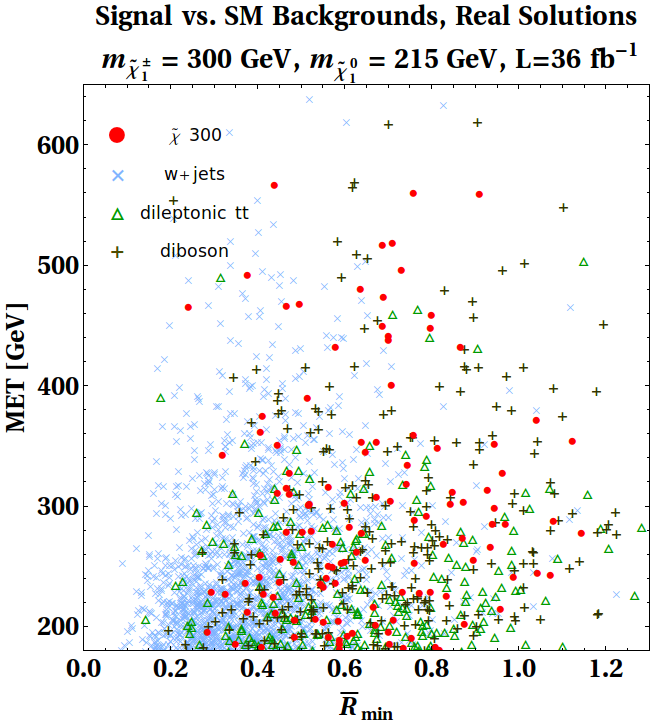}
\includegraphics[scale=0.35]{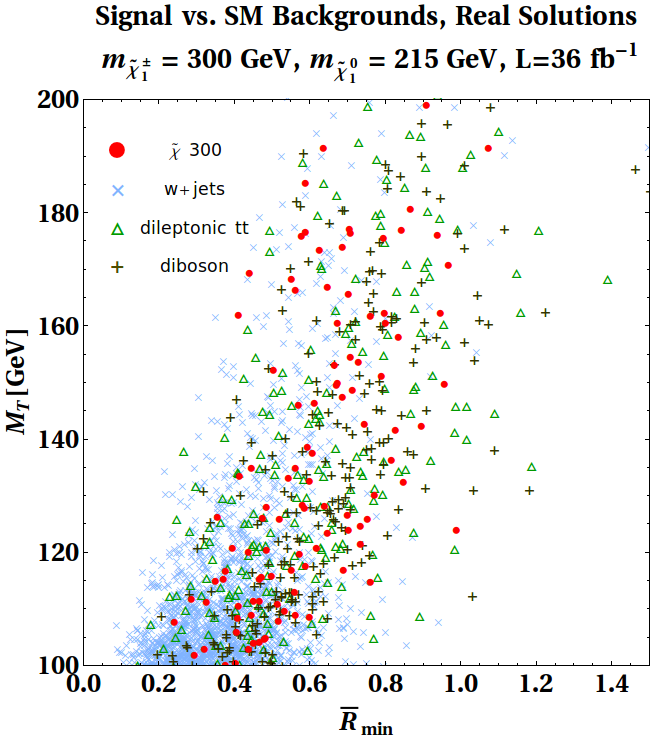}
\caption{Benchmark signals vs.\ various backgrounds, with normalized luminosity $L=36\, \ifb$. Left panel is for MET vs.\ $\Rmbmin$, Right panel is for $M_T$ vs.\ $\Rmbmin$}
\label{fig:MET-MT-Rmb}
\end{figure}

Due to the smaller $\Rmbmin$ values for the signal, one may want to lower the $\Rmbmin$ cut in the signal region selection. However, the backgrounds is large at lower $\Rmbmin$ values so a lower $\Rmbmin$ cut would need to be accompanied by a harder cut on other variables such as MET. The MET and $M_T$ vs.\ $\Rmbmin$ distributions for the signal and backgrounds are shown in Fig.~\ref{fig:MET-MT-Rmb}. We modify the two signal regions as follows:
\begin{itemize}

\item for events with $180 \leqslant$ MET $<350$~GeV, $M_T \geqslant 150$~GeV and $\Rmbmin \geqslant 0.7$ are required (SRL);

\item for events with MET $\geqslant350$~GeV, $M_T \geqslant 120$~GeV and $\Rmbmin \geqslant 0.4$ are required (SRH).
\end{itemize}

\begin{table}[ht]
\captionsetup{singlelinecheck = false, format= hang, justification=raggedright, font=footnotesize, labelsep=space}
\begin{small}
\begin{tabular}{l|ccccccc}
             & Initial & Prelim &  SRL(R) & SRH(R) &SRL(C) & SRH(C) \\ 
\hline\hline
$\tilde{\chi}^{\pm}\tilde{\chi}^{\mp}300$ & $6.84 \times 10^3$ & $5.83\times 10^1$ & 6.7 & 6.2 & 5.9 & 2.4 \\

$\tilde{\chi}^{\pm}\tilde{\chi}^0300$ & $1.37 \times 10^4$ & $1.20\times 10^2$ & 14.1 & 13.5 & 14.5 & 7.1 \\
\hline
$\tilde{\chi}^{\pm}\tilde{\chi}^{\mp}450$ & $1.27 \times 10^3$ & $2.17\times 10^1$ & 3.3 & 2.6 & 2.1 & 1.7 \\

$\tilde{\chi}^{\pm}\tilde{\chi}^0450$ & $2.64 \times 10^3$ & $9.24\times 10^1$ & 6.4 & 6.6 & 8.6 & 3.5 \\

\hline
\hline
$W$+ jets & $1.94\times 10^6$ & $2.09\times 10^3$& 30.2 &27.9&34.1 & 9.3 \\

di-boson & $4.37 \times 10^5$ &  $5.03\times 10^2$ & 54.2 &35.5 & 119 & 31.1 \\

$tt$ (Dilep) &$ 1.97\times 10^6$ & $4.78\times 10^2$ &60.1 & 6.6 &96.1 & 8.7 \\

$tt$ (Semilep) & 7.86 $\times 10^6$ & $1.82\times 10^1$ & 0 & 0 & 0 & 0\\

SM others & $3.50 \times 10^{6}$ &$1.22\times10^{2}$ & 15.5 & 7.1 & 20.1 & 0.9\\
\hline\hline
SM total & $1.58\times 10^7$  & $3.21 \times 10^3$ &  160 & 77.5 & 269 & 50.0 \\
\end{tabular} 
\end{small}
\caption{The cut flow for the chargino/neutrino analysis with 300 and 450~GeV benchmark signal and major SM backgrounds, assuming an integrated luminosity of 36  fb$^{-1}$.}
\label{tab:chargino_flow}
\end{table}

The numbers of events passing the above cuts are listed in Table~\ref{tab:chargino_flow}. In the table we separate the signal events into the chargino pair production and the chargino-neutralino production, and also events with real and complex solutions for $\Rmb$. 
If we include the $10\%$ background uncertainty, the signal significance of the 300~GeV chargino benchmark would be around 2.3 including all channels. In other words, this mass point can be excluded if no excess is present in the current data.  This exceeds the current multi-lepton bound on $m_{\tilde{\chi}_1^{\pm},  \tilde{\chi}_2^0}$ in the $W$ corridor, which is no higher than $250$~GeV~\cite{Sirunyan:2017lae,CMS:2017sqn,ATLAS:2017uun}. 

To explore the potential power of this channel in the future, we project the analyses for several mass points along the $W$ corridor to a higher integrated luminosity of $L= 300~\ifb$. Numerical results are presented in Table~\ref{tab:chargino300}, also with their overall signal efficiency. It is clear that as the chargino becomes heavier, the expectation of $\Rmb$ also increases, rendering a larger MET and $\Rmbmin$ on average. As the result, the overall signal efficiency grows and partially compensates the reduction of the production cross section. From the results we can see that for $300~\ifb$ 13~TeV LHC the reach in the chargino mass around the $W$ corridor could go beyond 400 GeV in the one lepton channel based on this method.
\begin{table}
\captionsetup{singlelinecheck = false, format= hang, justification=raggedright, font=footnotesize, labelsep=space}
\begin{tabular}{c|cccc}
$m_{\tilde{\chi}}$ (GeV) & 300 & 350 & 400 & 450 \\ 
\hline 
\hline 
Overall signal efficiency ($\epsilon_{\rm{tot}}$) & 3.4$\times 10^{-3}$ & 6.1$\times 10^{-3}$ & 7.9$\times 10^{-3}$ & 8.8$\times 10^{-3}$ \\  
Significance ($\sigma$) & 3.2 & 3.0 & 2.5 & 1.6 \\ 
\end{tabular} 
\caption{The projected overall signal efficiency $\epsilon_{\rm{tot}}$ and significance with $L=300\, \ifb$, in the presence of $10\%$ systematic background uncertainty. The increase in efficiencies for higher masses slows down the decrease in significances due to the smaller production cross sections. The $\epsilon_{\rm{tot}}$ is defined by the signal efficiency from summing all solutions and signal regions.}
\label{tab:chargino300}
\end{table}

\section{Conclusions}
\label{sec:conclusions}

SUSY searches at the LHC have put strong bounds on the superpartner masses, generally beyond 1 TeV for colored states. However, it is important to cover the search holes at lower mass regions before one can declare that SUSY is too heavy to address the naturalness problem. The search holes at lower masses arise when the superpartners have a compressed spectrum. The visible particles and the missing transverse energy from the cascade decays of the superpartner are soft in this case. They are often difficult to be distinguished from the SM backgrounds, which results in the weaker limits. To improve the searches, a useful strategy is to consider the superpartner production together with a hard ISR jet. The LSPs recoiled against the ISR will produce a larger missing transverse energy, allowing better identification of the SUSY signals.

In the case of stop search, the $R_M$ (or $\Rmb$) variable that measures the ratio of the LSP mass and the stop mass has been shown to be a powerful discriminator for the signal and the backgrounds. It has been used to extend the exclusion reach to 590 GeV along the top corridor using the all hadronic channel. The semileptonic channel could have a similar reach, as the neutrino momentum can be solved from the kinematic constraints and hence its contribution to the MET can be subtracted to obtain the contribution from the LSPs.  It is then natural to explore whether similar methods can be applied to smaller mass differences between the stop and the LSP, such as in the $W$ corridor where the current experimental limits are still quite weak.

In this paper we extend the strategy to the stop search in semileptonic decays around the $W$ corridor. In this case the $b$-jets are too soft to be useful and we lose the kinematic constraint from an on-shell top intermediate state during the decay. As a result, the neutrino momentum can not be fully reconstructed and one can not obtain a unique $\Rmb$ value from the experimental measurements. Nevertheless, the rest of the kinematic constraints still impose a restriction on the allowed $\Rmb$ values consistent with an event. We found that the minimum value of $\Rmb$ of the allowed interval provides a good discriminator for the signal and the backgrounds. By combining with other standard variables like MET and $M_T$, it can significantly extend the exclusion reach beyond the limits obtained from the current experimental analyses. The same analysis also applies to chargino/neutralino search in the $W$ corridor. The search reach of the chargino mass is not as good as the stop mass due to the smaller production cross section, but can still surpass the limits set by the current multilepton searches.

A lesson from these studies is that by fully utilizing the kinematic features and constraints of the signal and background events, one can construct discriminating variables that more effectively separate them, and therefore improve the search coverage. This is important for the difficult parameter regions where the signal and backgrounds have similar distributions in simple traditional variables.  The stake of finding or excluding new physics is so high that no stone should be left unturned. Coming up with better search strategies at the LHC shall continue to be a high priority in high energy physics.

\section*{Acknowledgments}
This work is supported in part by the US Department of Energy grant DE-SC-000999. H.-C.~C. was also supported by The Ambrose Monell Foundation at the Institute for Advanced Study, Princeton.

\appendix
\section{Solving for $\bar{R}_{\text{min}}$ and $\bar{R}_{\text{max}}$}
\label{app:technical}
In this Appendix, we explain in details of how we calculate the allowed $\Rmb$ range from kinematic constraints and present the analytical formula for $\bar{R}_{\text{min}}$ and $\bar{R}_{\text{max}}$. 

Assuming that the missing transverse momentum comes purely from the two neutralinos and the neutrino, from Eq.~(\ref{eq:Rmb}), we can write the neutrino's transverse momentum in terms of $\Rmb,\, \slashed{p}_T$ and the transverse momentum of ISR: $ \vec{p}_{T(J_{\text{ISR}})}\equiv (p_{jx}, ~p_{jy}) $:
\begin{equation}\label{eqa2}
\begin{split}
p_{\nu x}=\slashed{p}_x+\Rmb\times p_{jx}, \\
p_{\nu y}=\slashed{p}_y+\Rmb\times p_{jy} .
\end{split}
\end{equation}
After substituting them into the mass shell equations, (\ref{eq:mass_shell}), (\ref{eq:mass_shell2}),
\begin{equation}\label{eqa1}
\begin{split}
p_\nu^2 &= E^2_\nu-p^2_{\nu x}-p^2_{\nu y}-p^2_{\nu z}= 0, \\
(p_\ell+p_\nu)^2 = 2p_\nu \cdot p_\ell &= 2(E_\nu E_\ell-p_{\nu x}p_{\ell x}-p_{\nu y}p_{\ell y}-p_{\nu z} p_{\ell z})=m_{W}^2,
\end{split}
\end{equation}
we obtain a quadratic equation of $p_{\nu z}$ in the form of
\begin{equation}
ap^2_{\nu z}+bp_{\nu z}+c=0, 
\end{equation}
where coefficients $a,b,c$ are functions of $\Rmb$ given below:
\begin{flalign}\label{eqa3}
\begin{aligned}
 a =& ~E^2_\ell-p^2_{\ell z},\\
 b =& -(2p_{\ell x} p_{\ell z} p_{jx}\Rmb+2p_{\ell x}p_{\ell z}\slashed{p}_x+2p_{\ell y}p_{\ell z}p_{jy}\Rmb+2p_{\ell y}p_{\ell z}\slashed{p}_y+m^2_Wp_{\ell z}), \\
 c =& ~(E^2_\ell-p^2_{\ell x})p^2_{jx}\Rmb^2+2(E^2_\ell-p^2_{\ell x})p_{jx}\slashed{p}_x\Rmb+(E^2_\ell-p^2_{\ell x})\slashed{p}^2_x\\
&+(E^2_\ell-p^2_{\ell y})p^2_{jy}\Rmb^2+2(E^2_\ell-p^2_{\ell y})p_{jy}\slashed{p}_y\Rmb+(E^2_\ell-p^2_{\ell y})\slashed{p}^2_y\\
&-2p_{\ell x}p_{\ell y}[p_{jx}p_{jy}\Rmb^2+(p_{jx}\slashed{p}_y+p_{jy}\slashed{p}_x)\Rmb+\slashed{p}_x\slashed{p}_y]-m^2_Wp_{\ell x}p_{jx}\Rmb\\
&-m^2_Wp_{\ell x}\slashed{p}_x-m^2_Wp_{\ell y}p_{jy}\Rmb-m^2_Wp_{\ell y}\slashed{p}_y-\frac{m^4_W}{4} .\\
\end{aligned}
\end{flalign}

In order for the event process to be physical, the neutrino's momentum must be real which means coefficients $a,b,c$ satisfy the following inequality
\begin{equation}\label{eqa4}
b^2-4ac\geqslant0 .
\end{equation}
From Eq.~(\ref{eqa3}) we can see that $b$ is a linear function of $\Rmb$, $c$ is a quadratic function of $\Rmb$ and $a$ is a constant. Thus, the left part of inequality~ (\ref{eqa4}) is actually a quadratic function of $\Rmb$ and can be written in the form
\begin{equation}\label{eqa5}
b^2-4ac\geqslant0 \qquad \Rightarrow\qquad A\Rmb^2+B\Rmb+C\geqslant0,
\end{equation}
with the coefficients $A,B,C$ given by
\begin{flalign}\label{eqa6}
\begin{aligned}
A=& ~4p^2_{\ell z}(p_{\ell x}p_{jx}+p_{\ell y}p_{jy})^2,
-4(E^2_\ell-p^2_{\ell z})[(E^2_\ell-p^2_{\ell x})p^2_{jx}+(E^2_\ell-p^2_{\ell y})p^2_{jy}-2p_{\ell x}p_{\ell y}p_{jx}p_{jy}] \\
B=& ~4p^2_{\ell z}(p_{\ell x}p_{jx}+p_{\ell y}p_{jy})(2p_{\ell x}\slashed{p}_x+2p_{\ell y}\slashed{p}_y+m^2_W)+4(E^2_\ell-p^2_{\ell z})(m^2_Wp_{\ell x}p_{jx}+m^2_Wp_{\ell y}p_{jy})\\
& -4(E^2_\ell-p^2_{\ell z})[2(E^2_\ell-p^2_{\ell x})p_{jx}\slashed{p}_x+2(E^2_\ell-p^2_{\ell y})p_{jy}\slashed{p}_y-2p_{\ell x}p_{\ell y}(p_{jx}\slashed{p}_y+p_{jy}\slashed{p}_x)],\\
C=& -4(E^2_\ell-p^2_{\ell z})[(E^2_\ell-p^2_{\ell x})\slashed{p}^2_x+(E^2_\ell-p^2_{\ell y})\slashed{p}^2_y-2p_{\ell x}p_{\ell y}\slashed{p}_x\slashed{p}_y-m^2_Wp_{\ell x}\slashed{p}_x-m^2_Wp_{\ell y}\slashed{p}_y-\frac{m^4_W}{4}]\\
& +(2p_{\ell x}p_{\ell z}\slashed{p}_x+2p_{\ell y}p_{\ell z}\slashed{p}_y+m^2_Wp_{\ell z})^2 .\\
\end{aligned}
\end{flalign}

One can show that the coefficient $A$ is negative as long as $p_{\ell x}p_{jy}\not=p_{\ell y}p_{jx}$  by the following equivalent relations:
\begin{equation}\label{eqa7}
\begin{gathered}
(p^2_{\ell x}+p^2_{\ell y})[(p^2_{\ell y}+p^2_{\ell z})p^2_{jx}+(p^2_{\ell x}+p^2_{\ell z})p^2_{jy}]>p^2_{\ell z}(p^2_{\ell x}p^2_{jx}+p^2_{\ell y}p^2_{jy}+2p_{\ell x}p_{\ell y}p_{jx}p_{jy})\\
+2(p^2_{\ell x}+p^2_{\ell y})p_{\ell x}p_{\ell y}p_{jx}p_{jy}\\
\Downarrow\\
p^2_{\ell x}p^2_{\ell y}p^2_{jx}+p^2_{\ell x}(p^2_{\ell x}+p^2_{\ell z})p^2_{jy}+p^2_{\ell y}(p^2_{\ell y}+p^2_{\ell z})p^2_{jx}+p^2_{\ell x}p^2_{\ell y}p^2_{jy}\\
>2(p^2_{\ell x}+p^2_{\ell y}+p^2_{\ell z})p_{\ell x}p_{\ell y}p_{jx}p_{jy}\\
\Downarrow\\
(p^2_{\ell x}+p^2_{\ell y}+p^2_{\ell z})(p^2_{\ell x}p^2_{jy}+p^2_{\ell y}p^2_{jx})>2(p^2_{\ell x}+p^2_{\ell y}+p^2_{\ell z})p_{\ell x}p_{\ell y}p_{jx}p_{jy}\\
\end{gathered}
\end{equation}

The coefficients $A,B,C$ can be calculated from the experimentally measured lepton momentum $p_\ell$, transverse momentum of the ISR, $ \vec{p}_{T(J_{\text{ISR}})}=(p_{jx}, ~p_{jy}) $, and the missing transverse momentum $\slashed{p}_T$ for each event. After calculating  $A,B,C$, we can obtain the allowed $\Rmb$ range which must satisfy inequality~(\ref{eqa5}):
\begin{itemize}
\item if $B^2-4AC\geqslant0$, $\Rmb\in [\frac{-B+\sqrt{B^2-4AC}}{2A},\frac{-B-\sqrt{B^2-4AC}}{2A}]$;
\item if $B^2-4AC<0$, $\Rmb$ has no real solutions.
\end{itemize}
The second case may be caused by experimental smearing effects, a wrongly identified ISR system, or even wrong topologies in the case of backgrounds. Instead of simply discarding these events, in our study we also perform an analysis of these events by taking the real part of the solution and define $\Rmb (=\Rmbmin=\Rmbmax)\equiv-\frac{B}{2A}$ in case that the no real solution result is due to the experimental smearing and the real part may be close to the true $\Rmb$ value.

\section{Validating the Usefulness of the $\bar{R}_M$ Variables}
\label{app:validation}

To verify the usefulness of $\bar{R}_M$ for the semileptonic stop search in the $W$ corridor, we compare the significance of an analysis mainly using $M_T$, ISR and MET with that of the analysis including the $\Rmb$ variables. We use the benchmark Stop-450 for the numerical study. 

Both signal and background events need to satisfy preliminary selection rules as mentioned in Sec.~\ref{sec:Bench}. For the ``control'' study without $\Rmb$ variables, we optimize our selection cut so that the number of signal events $N_{\rm sig}$ is similar to that in our stop search with $\Rmb$ variables while minimizing the number of background events in order to have a fair comparison of the two analyses. To be specific, we require $M_T\geqslant130~\text{GeV},~\slashed{p}_T\geqslant280~\text{GeV},~~\text{and}~\slashed{p}_T\geqslant\frac{5}{7}\times p_{T(J_{\text{ISR}})}+28$~GeV. After applying this cut, we get 86.8 signal events and 247.8 background events for 36 $\ifb$ luminosity. 

The significance is calculated with a 10\% independent background systematic uncertainty using Eq.~(\ref{likelihood}), and the ``control'' study without $\Rmb$ variables gives a significance $\sim 2.9$. Compared to the significance $\sim 4.3$ from the analysis including the $\Rmbmin$ variable done in Sec.~\ref{sec:Stop450}, the $\Rmbmin$ variable provides a significant improvement. The extent of the improvement depends on the benchmark points and the choices of the signal regions, but the usefulness of the $\Rmb$ variables is very general.

\appendix


\begin{thebibliography}{10}

\bibitem{Aaboud:2017ayj} 
  M.~Aaboud {\it et al.} [ATLAS Collaboration],
  ``Search for a scalar partner of the top quark in the jets plus missing transverse momentum final state at $\sqrt{s}$=13 TeV with the ATLAS detector,''
  arXiv:1709.04183 [hep-ex].
  

\bibitem{Aaboud:2017bac} 
  M.~Aaboud {\it et al.} [ATLAS Collaboration],
  ``Search for squarks and gluinos in events with an isolated lepton, jets and missing transverse momentum at $\sqrt{s}$ = 13 TeV with the ATLAS detector,''
  arXiv:1708.08232 [hep-ex].

\bibitem{Aaboud:2017nfd} 
  M.~Aaboud {\it et al.} [ATLAS Collaboration],
  ``Search for direct top squark pair production in final states with two leptons in $\sqrt{s} = 13$ TeV $pp$ collisions with the ATLAS detector,''
  arXiv:1708.03247 [hep-ex].

\bibitem{Sirunyan:2017cwe} 
  A.~M.~Sirunyan {\it et al.} [CMS Collaboration],
  ``Search for supersymmetry in multijet events with missing transverse momentum in proton-proton collisions at 13 TeV,''
  Phys.\ Rev.\ D {\bf 96}, no. 3, 032003 (2017)
  doi:10.1103/PhysRevD.96.032003
  [arXiv:1704.07781 [hep-ex]].

\bibitem{Sirunyan:2017wif} 
  A.~M.~Sirunyan {\it et al.} [CMS Collaboration],
  ``Search for direct production of supersymmetric partners of the top quark in the all-jets final state in proton-proton collisions at $ \sqrt{s}=13 $ TeV,''
  JHEP {\bf 1710}, 005 (2017)
  doi:10.1007/JHEP10(2017)005
  [arXiv:1707.03316 [hep-ex]].
  
\bibitem{Sirunyan:2017kqq} 
  A.~M.~Sirunyan {\it et al.} [CMS Collaboration],
  ``Search for new phenomena with the MT2 variable in the all-hadronic final state produced in proton-proton collisions at sqrt(s) = 13 TeV,''
  arXiv:1705.04650 [hep-ex].


\bibitem{CMS:2017zki} 
  CMS Collaboration [CMS Collaboration],
  ``Search for supersymmetry using hadronic top quark tagging in 13 TeV pp collisions,''
  CMS-PAS-SUS-16-050.
  
\bibitem{Sirunyan:2017xse} 
  A.~M.~Sirunyan {\it et al.} [CMS Collaboration],
  ``Search for top squark pair production in pp collisions at $ \sqrt{s}=13 $ TeV using single lepton events,''
  JHEP {\bf 1710}, 019 (2017)
  doi:10.1007/JHEP10(2017)019
  [arXiv:1706.04402 [hep-ex]].
  
  
\bibitem{Sirunyan:2017leh} 
  A.~M.~Sirunyan {\it et al.} [CMS Collaboration],
  arXiv:1711.00752 [hep-ex].
  

\bibitem{Boehm:1999tr} 
  C.~Boehm, A.~Djouadi and Y.~Mambrini,
  ``Decays of the lightest top squark,''
  Phys.\ Rev.\ D {\bf 61}, 095006 (2000)
  doi:10.1103/PhysRevD.61.095006
  [hep-ph/9907428].

\bibitem{Das:2001kd} 
  S.~P.~Das, A.~Datta and M.~Guchait,
  ``Four-body decay of the stop squark at the upgraded Tevatron,''
  Phys.\ Rev.\ D {\bf 65}, 095006 (2002)
  doi:10.1103/PhysRevD.65.095006
  [hep-ph/0112182].

\bibitem{Konar:2016ata} 
  P.~Konar, T.~Mondal and A.~K.~Swain,
  ``Demystifying the compressed top squark region with kinematic variables,''
  Phys.\ Rev.\ D {\bf 96}, no. 9, 095011 (2017)
  doi:10.1103/PhysRevD.96.095011
  [arXiv:1612.03269 [hep-ph]].


\bibitem{CMS:2017odo} 
  CMS Collaboration [CMS Collaboration],
  ``Search for supersymmetry in events with at least one soft lepton, low jet multiplicity, and missing transverse momentum in proton-proton collisions at $\sqrt{s}=13~\mathrm{TeV}$,''
  CMS-PAS-SUS-16-052.
  
\bibitem{CMS:2016zvj} 
  CMS Collaboration [CMS Collaboration],
  ``Search for new physics in the compressed mass spectra scenario using events with two soft opposite-sign leptons and missing momentum energy at 13 TeV,''
  CMS-PAS-SUS-16-025.
 
\bibitem{Khachatryan:2016pxa} 
  V.~Khachatryan {\it et al.} [CMS Collaboration],
  ``Search for top squark pair production in compressed-mass-spectrum scenarios in proton-proton collisions at $\sqrt{s}$ = 8 TeV using the $\alpha_T$ variable,''
  Phys.\ Lett.\ B {\bf 767}, 403 (2017)
  doi:10.1016/j.physletb.2017.02.007
  [arXiv:1605.08993 [hep-ex]].
  
\bibitem{Hikasa:1987db} 
  K.~i.~Hikasa and M.~Kobayashi,
  ``Light Scalar Top at e+ e- Colliders,''
  Phys.\ Rev.\ D {\bf 36}, 724 (1987).
  doi:10.1103/PhysRevD.36.724

\bibitem{Muhlleitner:2011ww} 
  M.~Muhlleitner and E.~Popenda,
  ``Light Stop Decay in the MSSM with Minimal Flavour Violation,''
  JHEP {\bf 1104}, 095 (2011)
  doi:10.1007/JHEP04(2011)095
  [arXiv:1102.5712 [hep-ph]].
 
  
\bibitem{Sirunyan:2017kiw} 
  A.~M.~Sirunyan {\it et al.} [CMS Collaboration],
  ``Search for the pair production of third-generation squarks with two-body decays to a bottom or charm quark and a neutralino in proton-proton collisions at sqrt(s) = 13 TeV,''
  arXiv:1707.07274 [hep-ex].
 
\bibitem{Aad:2014nra} 
  G.~Aad {\it et al.} [ATLAS Collaboration],
  ``Search for pair-produced third-generation squarks decaying via charm quarks or in compressed supersymmetric scenarios in $pp$ collisions at $\sqrt{s}=8~$TeV with the ATLAS detector,''
  Phys.\ Rev.\ D {\bf 90}, no. 5, 052008 (2014)
  doi:10.1103/PhysRevD.90.052008
  [arXiv:1407.0608 [hep-ex]].
     
\bibitem{Aaboud:2016tnv} 
  M.~Aaboud {\it et al.} [ATLAS Collaboration],
  ``Search for new phenomena in final states with an energetic jet and large missing transverse momentum in $pp$ collisions at $\sqrt{s}=13$??TeV using the ATLAS detector,''
  Phys.\ Rev.\ D {\bf 94}, no. 3, 032005 (2016)
  doi:10.1103/PhysRevD.94.032005
  [arXiv:1604.07773 [hep-ex]].
 
 \bibitem{Drees:2012dd} 
  M.~Drees, M.~Hanussek and J.~S.~Kim,
  ``Light Stop Searches at the LHC with Monojet Events,''
  Phys.\ Rev.\ D {\bf 86}, 035024 (2012)
  doi:10.1103/PhysRevD.86.035024
  [arXiv:1201.5714 [hep-ph]].


\bibitem{Carena:2008mj} 
  M.~Carena, A.~Freitas and C.~E.~M.~Wagner,
  ``Light Stop Searches at the LHC in Events with One Hard Photon or Jet and Missing Energy,''
  JHEP {\bf 0810}, 109 (2008)
  doi:10.1088/1126-6708/2008/10/109
  [arXiv:0808.2298 [hep-ph]].

\bibitem{Hagiwara:2013tva} 
  K.~Hagiwara and T.~Yamada,
  ``Equal-velocity scenario for hiding dark matter at the LHC,''
  Phys.\ Rev.\ D {\bf 91}, no. 9, 094007 (2015)
  doi:10.1103/PhysRevD.91.094007
  [arXiv:1307.1553 [hep-ph]].
  
\bibitem{An:2015uwa} 
  H.~An and L.~T.~Wang,
  ``Opening up the compressed region of top squark searches at 13 TeV LHC,''
  Phys.\ Rev.\ Lett.\  {\bf 115}, 181602 (2015)
  doi:10.1103/PhysRevLett.115.181602
  [arXiv:1506.00653 [hep-ph]].
  
\bibitem{Macaluso:2015wja} 
  S.~Macaluso, M.~Park, D.~Shih and B.~Tweedie,
  ``Revealing Compressed Stops Using High-Momentum Recoils,''
  JHEP {\bf 1603}, 151 (2016)
  doi:10.1007/JHEP03(2016)151
  [arXiv:1506.07885 [hep-ph]].
   

   
  
\bibitem{Jackson:2016mfb} 
  P.~Jackson, C.~Rogan and M.~Santoni,
  ``Sparticles in motion: Analyzing compressed SUSY scenarios with a new method of event reconstruction,''
  Phys.\ Rev.\ D {\bf 95}, no. 3, 035031 (2017)
  doi:10.1103/PhysRevD.95.035031
  [arXiv:1607.08307 [hep-ph]].
  
\bibitem{Cheng:2016mcw} 
  H.~C.~Cheng, C.~Gao, L.~Li and N.~A.~Neill,
  ``Stop Search in the Compressed Region via Semileptonic Decays,''
  JHEP {\bf 1605}, 036 (2016)
  doi:10.1007/JHEP05(2016)036
  [arXiv:1604.00007 [hep-ph]].

\bibitem{Cheng:2017dxe} 
  H.~C.~Cheng, C.~Gao and L.~Li,
  ``Compressed Stop Searches with Two Leptons and Two b-jets,''
  arXiv:1706.02805 [hep-ph].



\bibitem{Konar:2017oah} 
  P.~Konar, T.~Mondal and A.~K.~Swain,
  ``Constraining slepton and chargino through compressed top squark search,''
  arXiv:1710.08664 [hep-ph].


\bibitem{Sirunyan:2017lae} 
  A.~M.~Sirunyan {\it et al.} [CMS Collaboration],
  ``Search for electroweak production of charginos and neutralinos in multilepton final states in proton-proton collisions at $\sqrt{s}=$ 13 TeV,''
  arXiv:1709.05406 [hep-ex].
 
\bibitem{CMS:2017sqn} 
  CMS Collaboration [CMS Collaboration],
  ``Combined search for electroweak production of charginos and neutralinos in pp collisions at $\sqrt{s} = 13~\mathrm{TeV}$,''
  CMS-PAS-SUS-17-004.
  
  
\bibitem{Aad:2014nua} 
  G.~Aad {\it et al.} [ATLAS Collaboration],
  ``Search for direct production of charginos and neutralinos in events with three leptons and missing transverse momentum in $\sqrt{s} =$ 8TeV $pp$ collisions with the ATLAS detector,''
  JHEP {\bf 1404}, 169 (2014)
  doi:10.1007/JHEP04(2014)169
  [arXiv:1402.7029 [hep-ex]].

\bibitem{Aad:2014vma} 
  G.~Aad {\it et al.} [ATLAS Collaboration],
  ``Search for direct production of charginos, neutralinos and sleptons in final states with two leptons and missing transverse momentum in $pp$ collisions at $\sqrt{s} =$ 8 TeV with the ATLAS detector,''
  JHEP {\bf 1405}, 071 (2014)
  doi:10.1007/JHEP05(2014)071
  [arXiv:1403.5294 [hep-ex]].


\bibitem{ATLAS:2017uun} 
  The ATLAS collaboration [ATLAS Collaboration],
  ``Search for electroweak production of supersymmetric particles in the two and three lepton final state at $\boldmath{\sqrt{s}=13\,}$TeV with the ATLAS detector,''
  ATLAS-CONF-2017-039.




\bibitem{Alwall:2014hca} 
  J.~Alwall {\it et al.},
  ``The automated computation of tree-level and next-to-leading order differential cross sections, and their matching to parton shower simulations,''
  JHEP {\bf 1407}, 079 (2014)
  doi:10.1007/JHEP07(2014)079
  [arXiv:1405.0301 [hep-ph]].
  

\bibitem{Sjostrand:2006za} 
  T.~Sjostrand, S.~Mrenna and P.~Z.~Skands,
  ``PYTHIA 6.4 Physics and Manual,''
  JHEP {\bf 0605}, 026 (2006)
  doi:10.1088/1126-6708/2006/05/026
  [hep-ph/0603175].


\bibitem{Mangano:2002ea} 
  M.~L.~Mangano, M.~Moretti, F.~Piccinini, R.~Pittau and A.~D.~Polosa,
  ``ALPGEN, a generator for hard multiparton processes in hadronic collisions,''
  JHEP {\bf 0307}, 001 (2003)
  doi:10.1088/1126-6708/2003/07/001
  [hep-ph/0206293].

\bibitem{deFavereau:2013fsa} 
  J.~de Favereau {\it et al.} [DELPHES 3 Collaboration],
  ``DELPHES 3, A modular framework for fast simulation of a generic collider experiment,''
  JHEP {\bf 1402}, 057 (2014)
  doi:10.1007/JHEP02(2014)057
  [arXiv:1307.6346 [hep-ex]].


\bibitem{Cacciari:2008gp} 
  M.~Cacciari, G.~P.~Salam and G.~Soyez,
  ``The Anti-k(t) jet clustering algorithm,''
  JHEP {\bf 0804}, 063 (2008)
  doi:10.1088/1126-6708/2008/04/063
  [arXiv:0802.1189 [hep-ph]].
    
\bibitem{Borschensky:2014cia} 
  C.~Borschensky, M.~Krämer, A.~Kulesza, M.~Mangano, S.~Padhi, T.~Plehn and X.~Portell,
  ``Squark and gluino production cross sections in pp collisions at $\sqrt{s}$ = 13, 14, 33 and 100 TeV,''
  Eur.\ Phys.\ J.\ C {\bf 74}, no. 12, 3174 (2014)
  doi:10.1140/epjc/s10052-014-3174-y
  [arXiv:1407.5066 [hep-ph]].

\bibitem{btagging}
  The ATLAS collaboration,
  ``Expected performance of the ATLAS b-tagging algorithms in Run-2,''
  ATL-PHYS-PUB-2015-022.


\bibitem{Sirunyan:2017mrs} 
  A.~M.~Sirunyan {\it et al.} [CMS Collaboration],
  ``Search for supersymmetry in events with one lepton and multiple jets exploiting the angular correlation between the lepton and the missing transverse momentum in proton-proton collisions at $\sqrt{s} = $ 13 TeV,''
  arXiv:1709.09814 [hep-ex].
  


\bibitem{Cowan:2010js} 
  G.~Cowan, K.~Cranmer, E.~Gross and O.~Vitells,
  ``Asymptotic formulae for likelihood-based tests of new physics,''
  Eur.\ Phys.\ J.\ C {\bf 71}, 1554 (2011)
  Erratum: [Eur.\ Phys.\ J.\ C {\bf 73}, 2501 (2013)]
  doi:10.1140/epjc/s10052-011-1554-0, 10.1140/epjc/s10052-013-2501-z
  [arXiv:1007.1727 [physics.data-an]].




\end{thebibliography}
\end{document}